\documentclass[aps, preprint]{revtex4-1}

\usepackage{dcolumn}
\usepackage{amsbsy}
\usepackage{amsmath}

\usepackage{amssymb}
\usepackage{braket}
\usepackage{color}
\usepackage{fullpage}
\usepackage{graphicx}
\usepackage{bm}

\usepackage{subfigure}
\usepackage[hidelinks]{hyperref}
\usepackage{epstopdf}

\usepackage{comment}
\graphicspath{{./Figures/}}

\newcommand{\be}{\bm{e}}

\newcommand{\bn}{\mathbf{n}}

\newcommand{\bu}{\bm{u}}

\newcommand{\bA}{\bm{A}}

\begin{document}
\title{Edge states and topological pumping in elastic lattices with periodically modulated coupling}
\author{Matheus I. N. Rosa$^{a}$, Raj Kumar Pal$^{b}$, Jos\'{e} R.F. Arruda$^c$ and  Massimo Ruzzene$^{a,b}$}
\affiliation{ $^a$ School of Mechanical Engineering, Georgia Institute of Technology, Atlanta GA 30332}
\affiliation{ $^b$ School of Aerospace Engineering, Georgia Institute of Technology, Atlanta GA 30332}
\affiliation{ $^c$ School of Mechanical Engineering, State University of Campinas (UNICAMP), Campinas, SP, Brasil}


\begin{abstract}
We investigate the dispersion topology of elastic lattices characterized by spatial stiffness modulation. The modulation is defined by the sampling of a two-dimensional surface, which provides the lattices with topological properties that are usually attributed to two-dimensional crystals. We show that the cyclic variation of the phase of the stiffness modulation leads to a Berry phase accumulation for the Bloch eigenmodes, which is characterized by integer valued Chern numbers for the bands and associated gap labels. The resulting non-trivial gaps are spanned by edge modes localized at one of the boundaries of the considered 1D lattices. The edge mode location is governed by the phase of the stiffness distribution, whose spatial modulation drives these modes to transition from one edge to the other, through a bulk state that occurs when the corresponding dispersion branch touches the bulk bands. This property enables the implementation of a topological pump that is obtained by stacking and coupling a family of modulated 1D lattices along a second spatial dimension. The gradual variation of the stiffness phase modulation drives the adiabatic transition of the edge states, which transition from their localized state at one boundary, to a bulk mode and, finally, to another localized state at the opposite boundary. Similar effects were illustrated under the assumption of quasiperiodic modulation of the lattice interactions. We here demonstrate that a topological pump can be achieved also in periodic media, and illustrate this for the first time in elastic discrete lattices.

\end{abstract}

\maketitle
\pagebreak

\section{Introduction}
The investigation of topologically protected modes is an area of significant interest for  elastic~\cite{mousavi2015topologically}, acoustic~\cite{yang2015topological,lu2017observation}, 
electromagnetic~\cite{lu2014topological} and quantum~\cite{hasan2010colloquium} systems. Quantum Hall effect~\cite{thouless1982quantized} analogues have recently received significant interest as part of the search for backscattering-free and defect-immune energy transport along boundaries or surfaces. The resulting robust dynamic properties and corresponding topological protection ensure that bulk bandgaps support wave modes that are localized and propagate along boundaries or interfaces. Two broad strategies have been pursued for the implementation of such topologically protected waves in classical systems. The first strategy encompasses implementations of quantum Hall effect (QHE) analogues that typically involve active elements that break time-reversal symmetry. Examples include rotating gyroscopes~\cite{wang2015topological,nash2015topological} and active fluids~\cite{souslov2017topological} arranged in either regular or irregular~\cite{mitchell2018amorphous} lattice networks. The second strategy involves emulating the quantum spin Hall effect (QSHE), with notable examples including coupled pendulums~\cite{susstrunk2015observation}, rotating disks~\cite{pal2016helical} and elastic media~\cite{mousavi2015topologically,chen2018self,chaunsali2018subwavelength}. Configurations belonging to the first category feature chiral edge states that propagate in one direction only. In contrast, QSHE analogues are characterized by helical edge modes of different polarization that travel in opposite directions, and that have the distinct advantage of achieving defect-immune energy transport using solely passive components. While initial QSHE analogues were obtained through rather intricate designs driven by the stringent requirements imposed by the desired topology of the dispersion branches, recent studies have exploited valley degrees of freedom~\cite{xiao2007valley} and dispersion branch doubling~\cite{lu2017observation} to considerably simplify the configurations. Examples include plates with stubs in hexagonal lattice arrangements~\cite{vila2017observation} and reconfigurable phononic elastic lattices that employ tunable prestrain to break spatial inversion symmetry~\cite{liu2018tunable}. Of note is the recent contribution by Barlas and Prodan~\cite{barlas2018topological} who have developed a set of transformations  which produce mechanical analogues of any corresponding quantum Hamiltonian. 

A recently explored approach involves accessing higher dimensional topological effects in lower dimensional physical systems~\cite{ozawa2016synthetic,kraus2016quasiperiodicity}. Notably, Zilberberg and coworkers achieved pumping of electromagnetic waves from one end to the other of quasi-periodic arrays of coupled waveguides~\cite{kraus2012topological,verbin2013observation}. Prodan and coworkers have demonstrated the presence of localized modes in a chain of mechanical spinners arranged in patterns obtained by following projection rules from selected manifolds~\cite{apigo2018topological}. A debate regarding the specific role of quasi-periodicity and incommensurability for achieving such topological effects has recently ensued~\cite{madsen2013topological,kraus2013comment}. We here investigate the topological properties of discrete 1D elastic lattices, where stiffnesses are periodically modulated in space by selecting their values through the sampling of a 2D surface. This leads to non-trivial topological properties associated with the higher dimensional super-space, which manifest themselves as localized modes at the boundaries of the 1D lattices. These modes occur in lattices obtained from periodic and quasi-periodic projections alike of the considered 2D surface. We also illustrate that a cyclic variation of the phase of the stiffness modulation leads to a Berry phase accumulation in the Bloch eigenvectors, which is quantified by integer valued Chern numbers, and is therefore related to the existence of robust topological modes~\cite{hatsugai1993chern}.  Bandgaps characterized by non-trivial labels are spanned by modes that are localized at either one of the two boundaries of the 1D lattice, depending on the phase of the stiffness modulation. When topologically protected modes touch the bulk bands, they transform from left-localized to right-localized, or viceversa. This property is employed to implement a topological pump, which is obtained by stacking the 1D lattices along a second direction along which the phase of the stiffness modulation is gradually varied. This causes the edge wave to adiabatically evolve from a localized state at one boundary, to a bulk wave, and finally to another localized state at the opposite boundary. While a similar adiabatic pump was demonstrated in quasiperiodic photonic media \cite{kraus2012topological}, here we show the same behavior for a periodic elastic system, and note that quasi-periodicity is not a strict requirement. Similar conclusions have been drawn for a family of trimer lattices, obtained from mapping onto the commensurate off-diagonal Aubry-Andr\'{e}-Haper model~\cite{harper1955single}, for which it is shown that the chiral edge modes have a topological origin inherited from this effective mapping~\cite{alvarez2018edge}.

\section{Background: 1D lattice with periodic stiffness modulation}\label{TheorySec}
\begin{figure}
	\centering
	\subfigure[]{\includegraphics[width=0.355\textwidth]{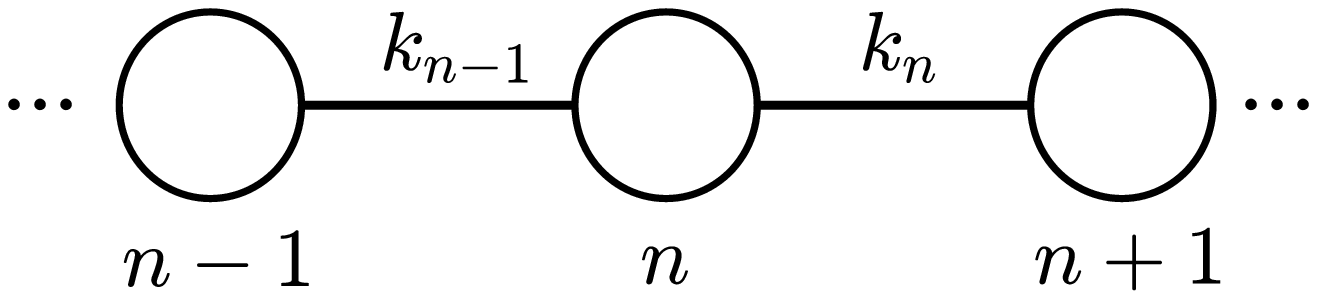}}
	\subfigure[]{\includegraphics[width=0.6\textwidth]{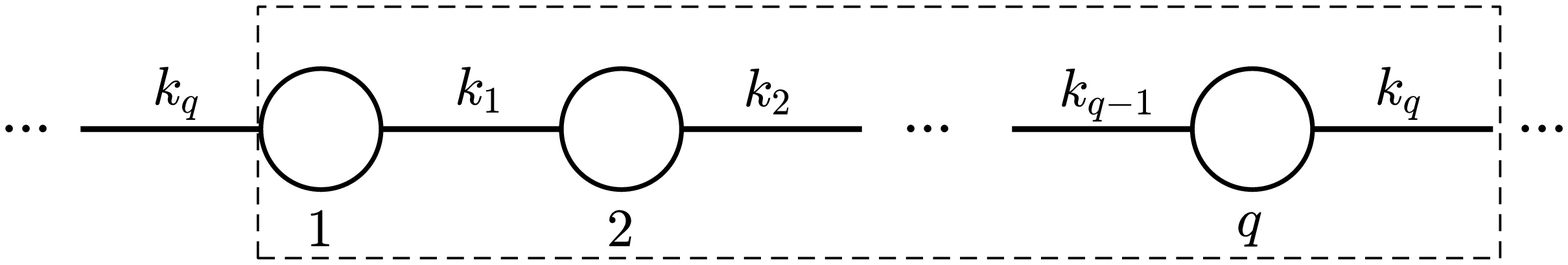}}
	\caption{1D discrete lattice with periodic stiffness modulation varying according to the law $k_n=k_0[1+\alpha\cos\left(2\pi pn/q + \phi\right)]$ (a), and schematic of a unit cell comprising $q$ masses (b).}
	\label{Fig:Chain1}
\end{figure}

We consider a spring-mass chain composed of equal masses $m$ (Fig. \ref{Fig:Chain1}a) and spring constants $k_n$ defined by an expression inspired by the Aubry-Andr\'{e} model for quasi-crystals~\cite{kraus2012topological}:
\begin{equation}\label{eqkn}
k_n = k_0\left[1 + \alpha\cos\left(2\pi n\dfrac{p}{q} + \phi \right)\right].
\end{equation}

\begin{figure}
	\centering
	\subfigure[]{\includegraphics[width=0.46\textwidth]{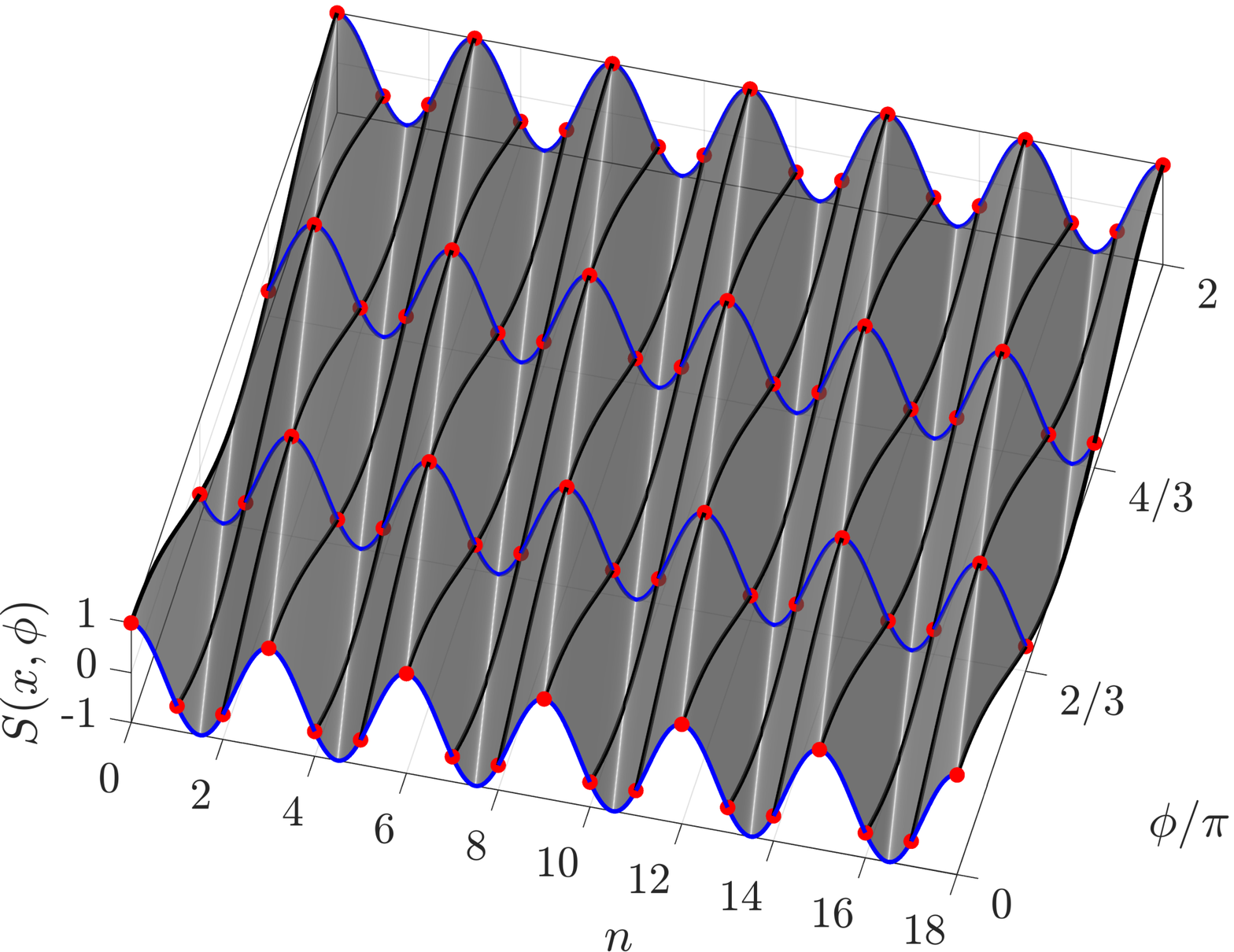}}
	\subfigure[]{\includegraphics[width=0.425\textwidth]{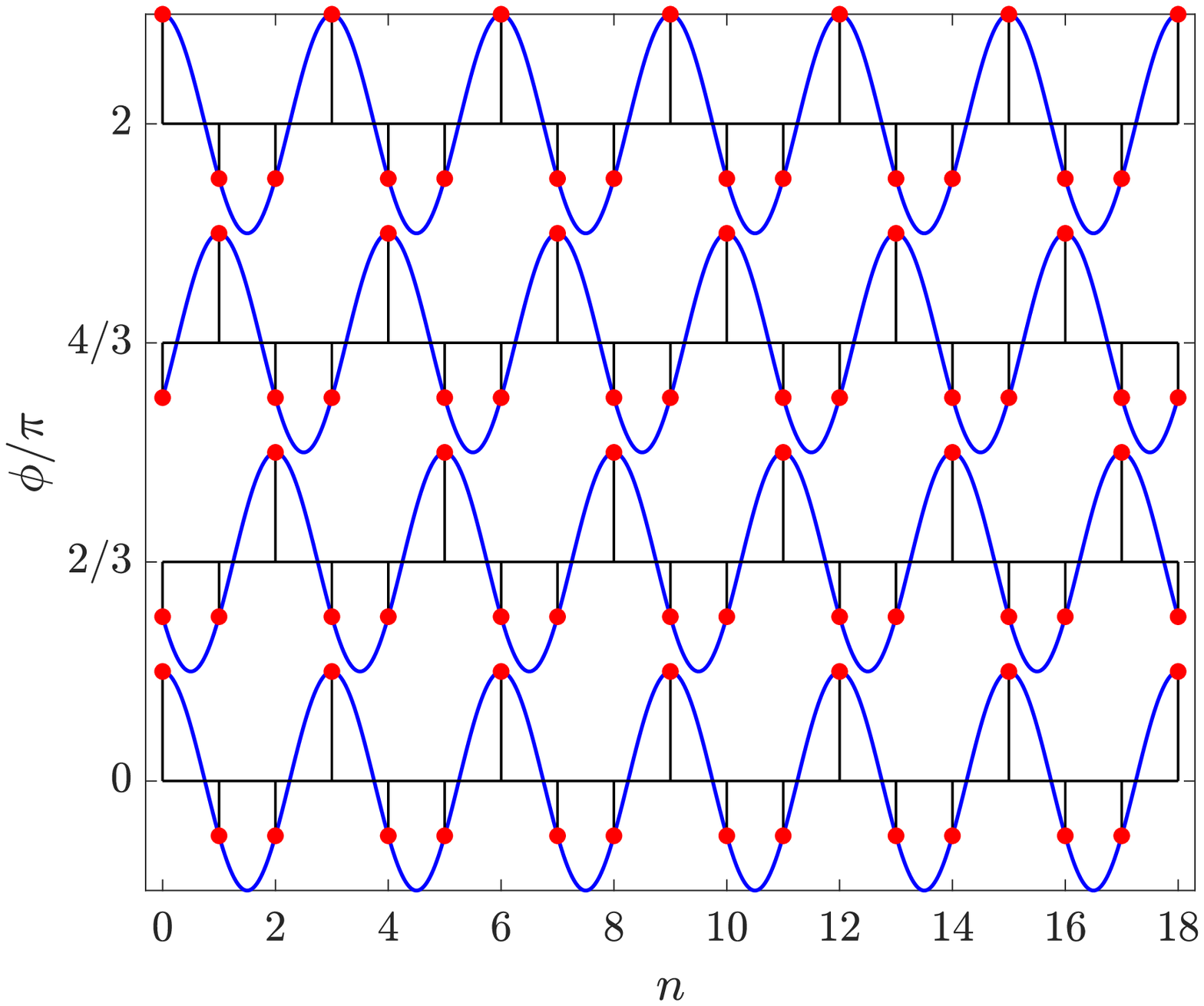}}
	\caption{Surface $\mathcal{S}(x,\phi)=\cos\left(2\pi \tau x + \phi\right)$ employed for generation of stiffness constants and sampled at $x_n=n$ and $\phi_k=(k-1) 2\pi/3$ ($k\in [1, \, 4]$) (red dots and blue lines) (a). Cross sections of the surface at $\phi_k=(k-1) 2\pi/3$ ($k\in [1, \, 4]$) showing the stiffness modulation as a function of phase and lattice location (b).}
	\label{Fig:Stiffnesses}
\end{figure}

Here $\alpha<1$, the ratio $p/q$ and $\phi$ respectively define amplitude, frequency and phase of the stiffness modulation. We consider the integers $p$ and $q$ to be co-prime such that $p/q$ is always an irreducible fraction. This enforces the spatial periodicity of the chain, which consists of irreducible unit cells of $q$ masses, \emph{i.e.} $k_{n+q}=k_n$ (Fig. \ref{Fig:Chain1}b). The stiffness modulation is the result of the sampling of a 2D surface $\mathcal{S}(x,\phi)=\cos\left(2\pi \tau x + \phi\right)$ at discrete values $x_n=n$, with the interatomic distance conveniently set equal to $1$ (Fig.~\ref{Fig:Stiffnesses}(a)). The lattice is thus the result of the sampling and projection in 1D of a higher dimensional space, here a 2D lattice. As such, the lattice retains the topological properties the 2D lattice that result from specific choices and variations of $\tau,\phi$. Of interest is specifically the lattice family associated with different values of $\phi$. For example, Fig.~\ref{Fig:Stiffnesses}b shows the spring constant variation along the lattice for $\tau=p/q=1/3$ and $\phi_k=(k-1) 2\pi/3$ ($k\in [1, \, 4]$). These discrete changes in $\phi_k$ produce a shift of the lattice by one position to the left, or by two positions to the right, and reproduce the initial lattice when $\phi_k=2 \pi$.

\subsection{Dispersion analysis and topology of the bulk bands}
The effect of the stiffness shifts on the lattice band structure and its topology are investigated next. The equation governing plane wave propagation at frequency $\omega$ may be written with reference to  mass $n$ as
\begin{equation}\label{govEqn2}
-m\omega^2u_n + (k_n+k_{n-1})u_n - k_nu_{n+1} - k_{n-1}u_{n-1}=0.
\end{equation}

\begin{figure}
	\centering
	\subfigure[]{\includegraphics[width=0.45\textwidth]{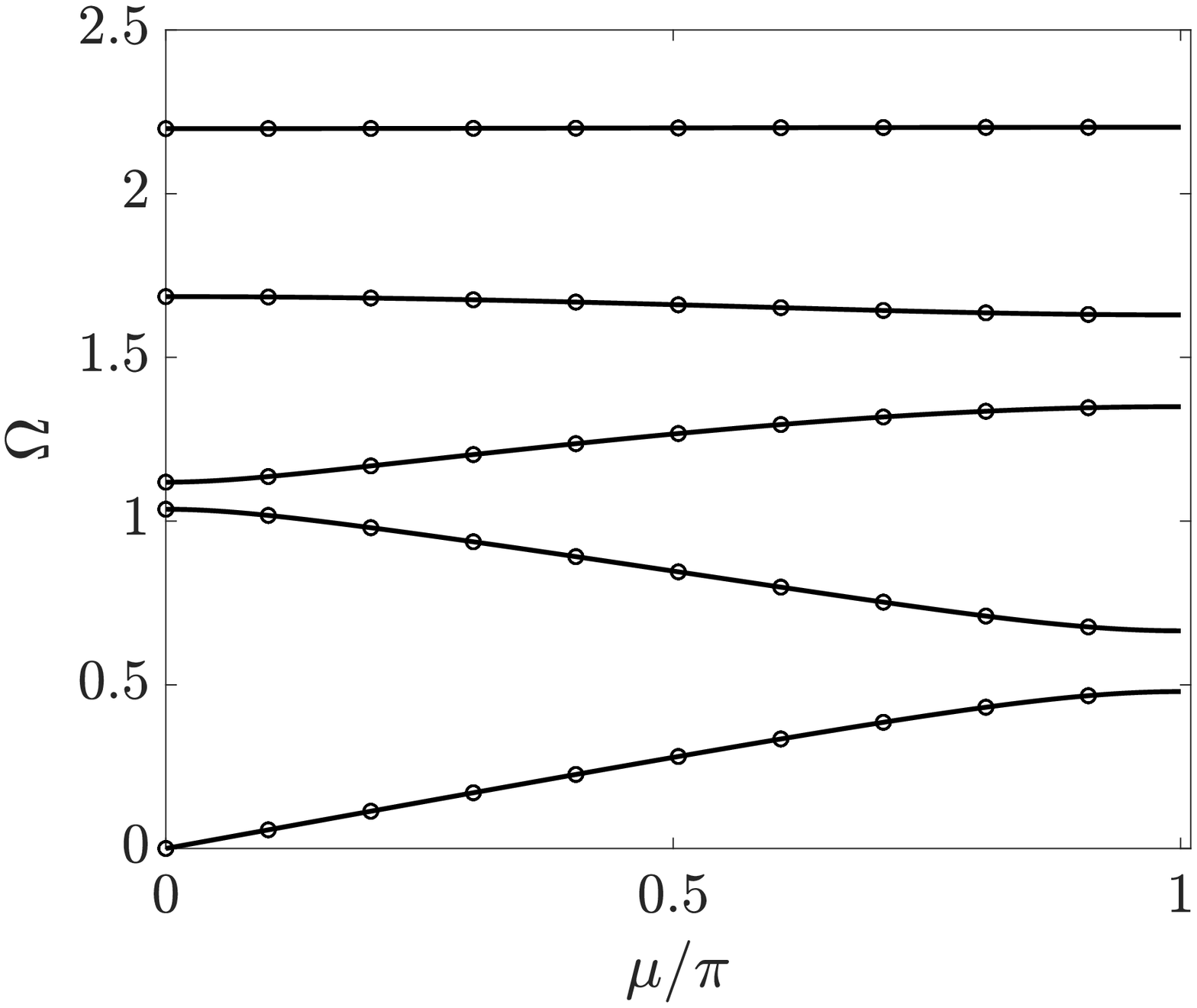}}
	\subfigure[]{\includegraphics[width=0.45\textwidth]{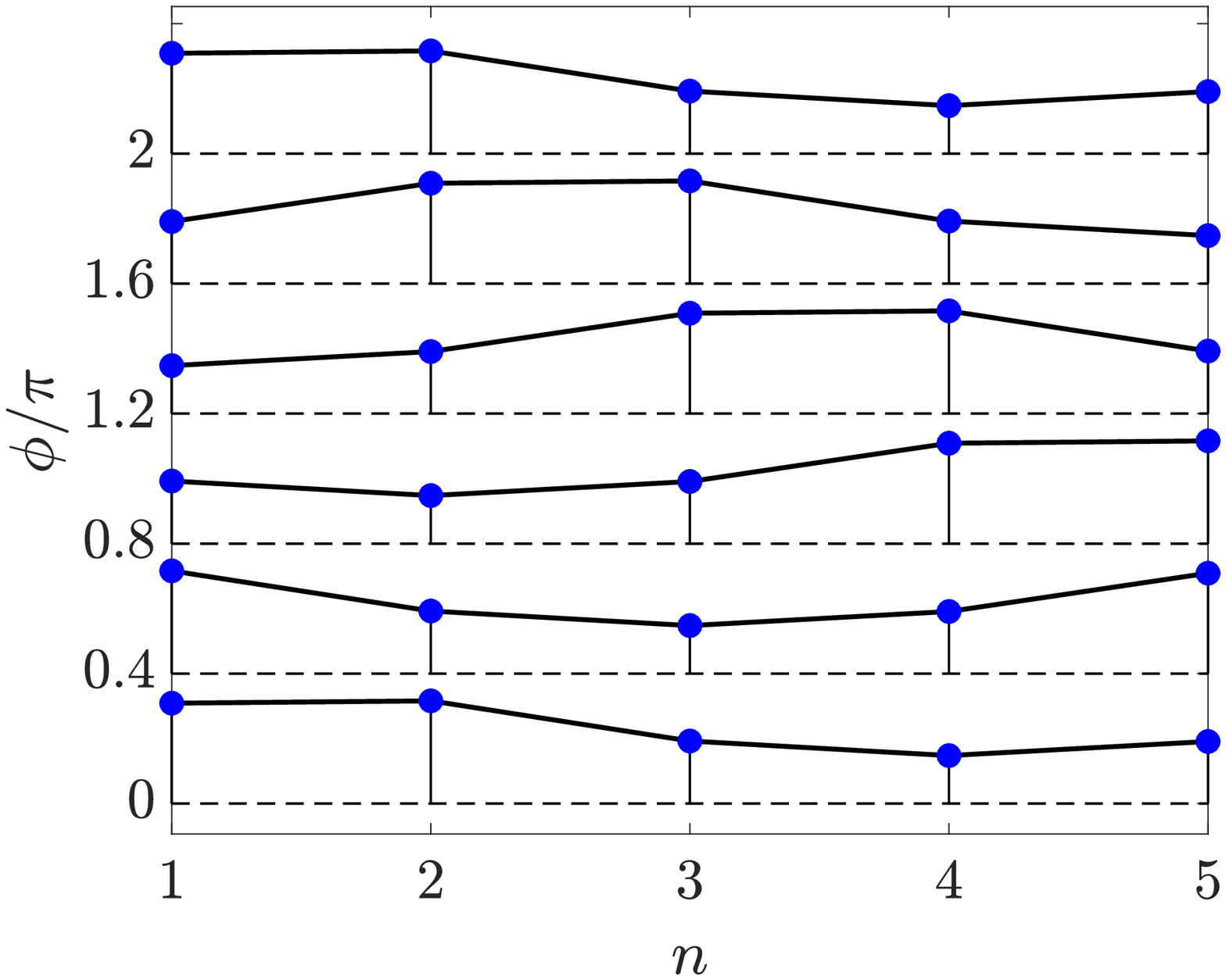}}
	\subfigure[]{\includegraphics[width=0.45\textwidth]{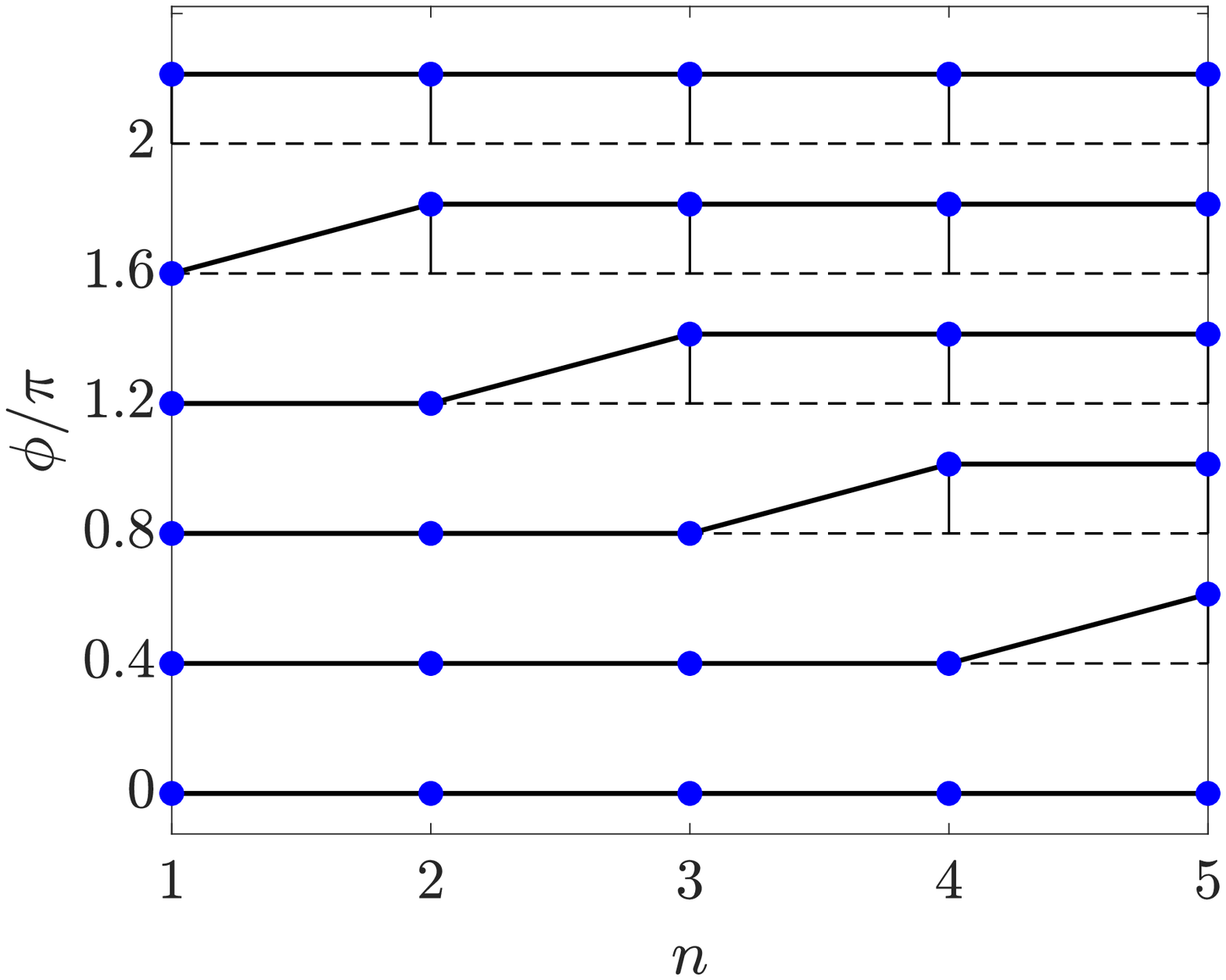}}
	\caption{Dispersion diagrams for the $p/q=1/5$ and $\alpha=0.6$ lattice showing $q=5$ branches and $q-1=4$ bandgaps. The diagram superimposes the curves for $\phi_k=0$ (solid line) and for $\phi_k=2/5 \pi$ (circles) showing that the eigenvalues are not affected by the phase shift in stiffness modulation (a). Variation of first Bloch mode evaluated at $\mu=0.6 \pi$ showing the amplitude shift corresponding to the stiffness modulation (b), and the additional phase accumulation that differentiates the mode for $\phi_k=0$ and $\phi_k=2\pi$ (c).}
	\label{Fig:Dispersion}
\end{figure}

Imposing Bloch periodicity conditions $u_{n+q}=e^{i\mu} u_n$, yields an eigenvalue problem in the form:
\begin{equation}\label{eq: eigenvalue 1}
\mathbf{K}(\mu)\mathbf{u}= \Omega^2 \mathbf{u},
\end{equation}where $\mathbf{K}(\mu)$ is a stiffness matrix and $\Omega = \omega/\omega_0$ is the non-dimensional frequency ($\omega_0=\sqrt{k_0/m}$). Solution for wavenumber $\mu \in [0,\pi]$ leads to eigenvalues that define the band structure of the lattice, along with the associated eigenvectors, or Bloch modes, $\mathbf{u}=(u_1, u_2, ... , u_q)^T$. The dispersion curves  for the representative case of a chain with $p/q=1/5$ and $\alpha=0.6$ are shown in Fig.~\ref{Fig:Dispersion}(a), where results for $\phi_k=0$ (solid lines) and $\phi_k=2/5 \pi$ (circles) are superimposed. As expected, the band structure includes $q$ bands and $q-1$ bandgaps~\cite{bellissard1982cantor}. The dispersion branches and the eigenvalue they represent are unaffected by the phase in stiffness modulation. However,  the phase shift manifests itself in a phase accumulation in the Bloch modes. Figures~\ref{Fig:Dispersion}(b,c) show the comparison of the first wave mode of the lattice computed for $\mu=0.6 \pi$ for $\phi_k=(k-1) 2\pi/5$ ($k\in [1, \, 6]$). Specifically, the modal amplitude $\mathbf{|u|}_q$ and its variation for the considered values of $\phi$ illustrates a left-shift by one unit, or a right shift in 4 units occurring for each increment in the phase angle, and show how the mode returns to its original magnitude distribution for $\phi_k=2 \pi$ (fig.~\ref{Fig:Dispersion}(b)). A phase accumulation also occurs in the Bloch modes, as illustrated by the spatial variation of the relative phase between eigenmodes corresponding to successive increments of  $\phi$ shown in in Fig.~\ref{Fig:Dispersion}(c). Such phase increment is obtained by aligning the first eigenmode for $\phi_1=0$ to the real axis through proper normalization, and subsequently apply it to the eigenvectors corresponding to the considered phase increments.

As Fig.~\ref{Fig:Stiffnesses} illustrates, a change in $\phi_k \to \phi_{k+1}$ produces a stiffness shift obtained by imposing that $k_n(\phi+2\pi/q) = k_{n+s}(\phi)$, where $s$ must obey the algebraic congruence relation:
\begin{equation}\label{PSeqn}
p s \equiv 1 \pmod{q}.
\end{equation}
Since $p$ and $q$ are coprime, Eqn.~\ref{PSeqn} identifies two solutions in the range $|s| < q$ \cite{burton2006elementary}, which correspond to the two possible shifts of the stiffness values along the lattice. For the case $p/q=1/3$ of Fig.~\ref{Fig:Stiffnesses}, the solutions of Eqn.~\eqref{PSeqn} are $s=1$ and $s=-2$, corresponding either to a shift of one position to the left or of two positions to the right, respectively. It can be easily found that for $p/q=1/5$ corresponding to Fig.~\ref{Fig:Dispersion}, the solutions of Eqn.~\eqref{PSeqn} are $s=1,-4$. As suggested by Fig.~\ref{Fig:Dispersion}, these shifts are reflected in the eigenvectors, or wave modes, of the lattice, whose components not only undergo a permutation of indexes according to $s$ defined above, but also accumulate a phase. In fact, by direct substitution of the stiffness values $k_n(\phi+2\pi/q) = k_{n+s}(\phi)$ in the Bloch eigenproblem it can be shown that the eigenvector $\bm{u}(\phi+2\pi/q)$ is related to $\bm{u}(\phi)$ as:
\begin{equation}\label{eqeigvectors}
    \bm{u} = (u_{s+1}, u_{s+2}, ... , u_q, e^{i\mu}u_1, ..., e^{i\mu}u_{s} )^T.
\end{equation}
where $s$ is the positive root of Eqn.~\eqref{PSeqn} in the range $|s|<q$. Hence, the components are shifted by $s$ units while the first $s$ components accumulate a phase of $e^{i\mu}$ as they are shifted to the tail of the eigenvector. It is easy to see that $\bm{u}$ can be multiplied by $e^{-i\mu}$ leading to:
\begin{equation}\label{eqeigvectors2}
    \bm{u}=(u_{s+1}e^{-i\mu}, u_{s+2}e^{-i\mu}, ... , u_q e^{-i\mu},u_1, ..., u_{s} )^T,
\end{equation}
where the last $q-s$ components accumulate a phase of $e^{-i\mu}$ as they are shifted to the beginning of the eigenvector. In both cases, variation of $\phi \in [0,\,\, 2\pi]$ in $q$ steps of $2\pi/q$ leads to
\begin{equation}\label{phaseDiff}
\bu(\mu,\phi + 2 \pi)  =  e^{i s \mu} \bu(\mu ,  \phi) . 
\end{equation}
where the total accumulated phase is given by the factor $e^{i s \mu}$.

The lattices considered herein can be considered as part of a family which is the result of the variation of the parameter $\phi$. In this context, the adiabatic cyclical evolution of this parameter leads to the accumulation of a Berry phase~\cite{nassar2018quantization}, here equal to $s \mu$. This phase is an observable topological invariant which is indicative of the nontrivial topology of the reciprocal space for the lattice family formed from the underlying 2D surface $\mathcal{S}(x,\phi)$. This invariant can be quantified by the Chern number associated to a given dispersion band defined in the $(\mu,\phi)\in \mathbb{T}^2 = [0,2\pi]\times [0,2\pi]$ space~\cite{thouless1982quantized,hatsugai1993chern}, which is given by
\begin{equation}\label{Cherneq1}
C = \dfrac{1}{2 \pi i} \int_{\mathcal{D}} \nabla \times \bA \; d\mathcal{D},
\end{equation}
where $\mathcal{D}$ is the base manifold, here $\mathbb{T}^2$, $\nabla = (\partial/\partial \mu)\be_{\mu} +  (\partial/\partial \phi)\be_{\phi} $
and $\bA = \bu^* \cdot \nabla \bu$, with $()^*$ denoting a complex conjugate. Applying Stoke's theorem to Eqn.~\eqref{Cherneq1} yields,  
\begin{equation}\label{ChernBoundary}
C = \dfrac{1}{2\pi i} \int_{\partial \mathcal{D}} \bn \times \bA \; d (\partial \mathcal{D}) .
\end{equation}

\begin{figure}
	\centering
	\subfigure[]{\includegraphics[width=0.45\textwidth]{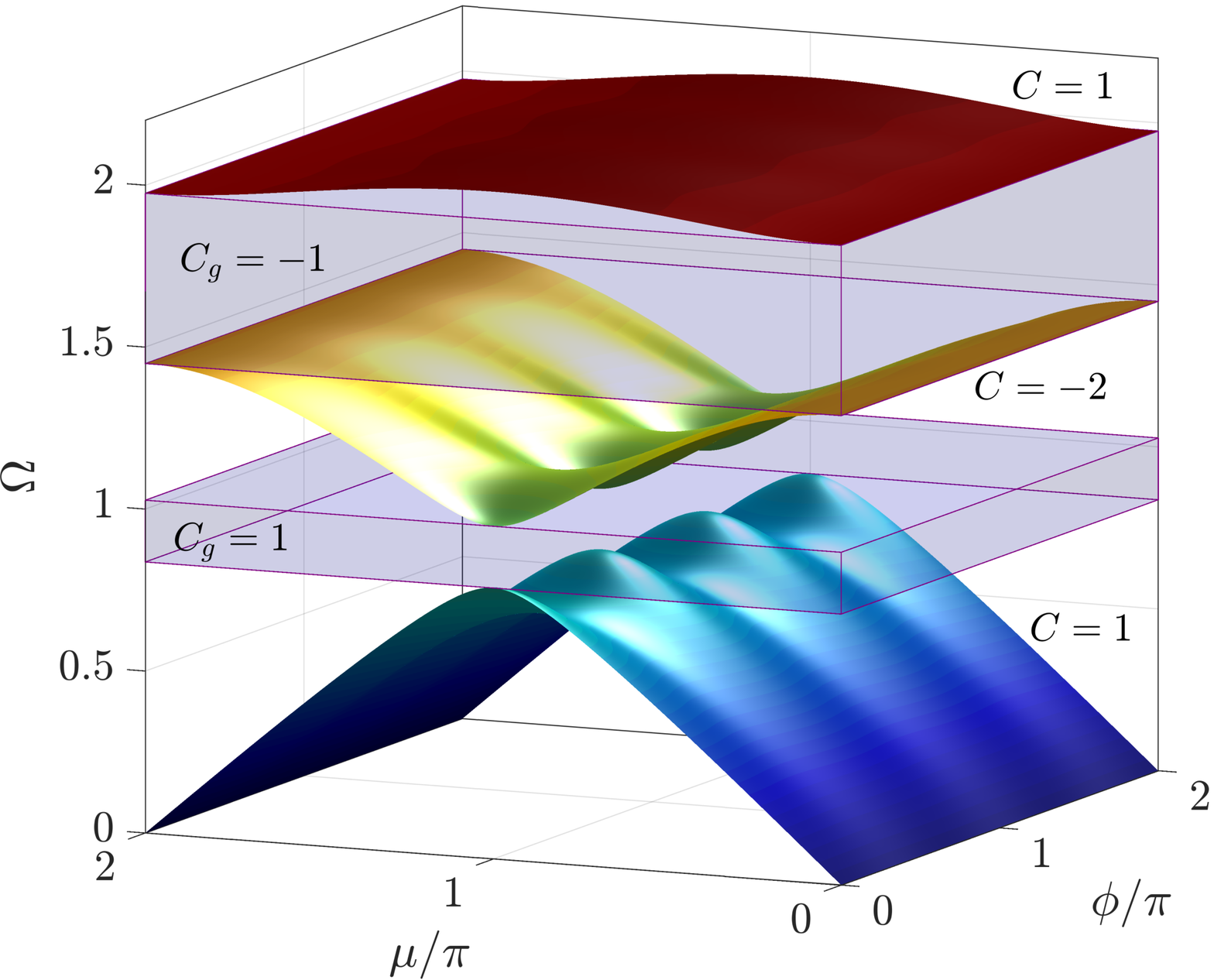}\label{Fig:Disp_unitcell}}
	\subfigure[]{\includegraphics[width=0.45\textwidth]{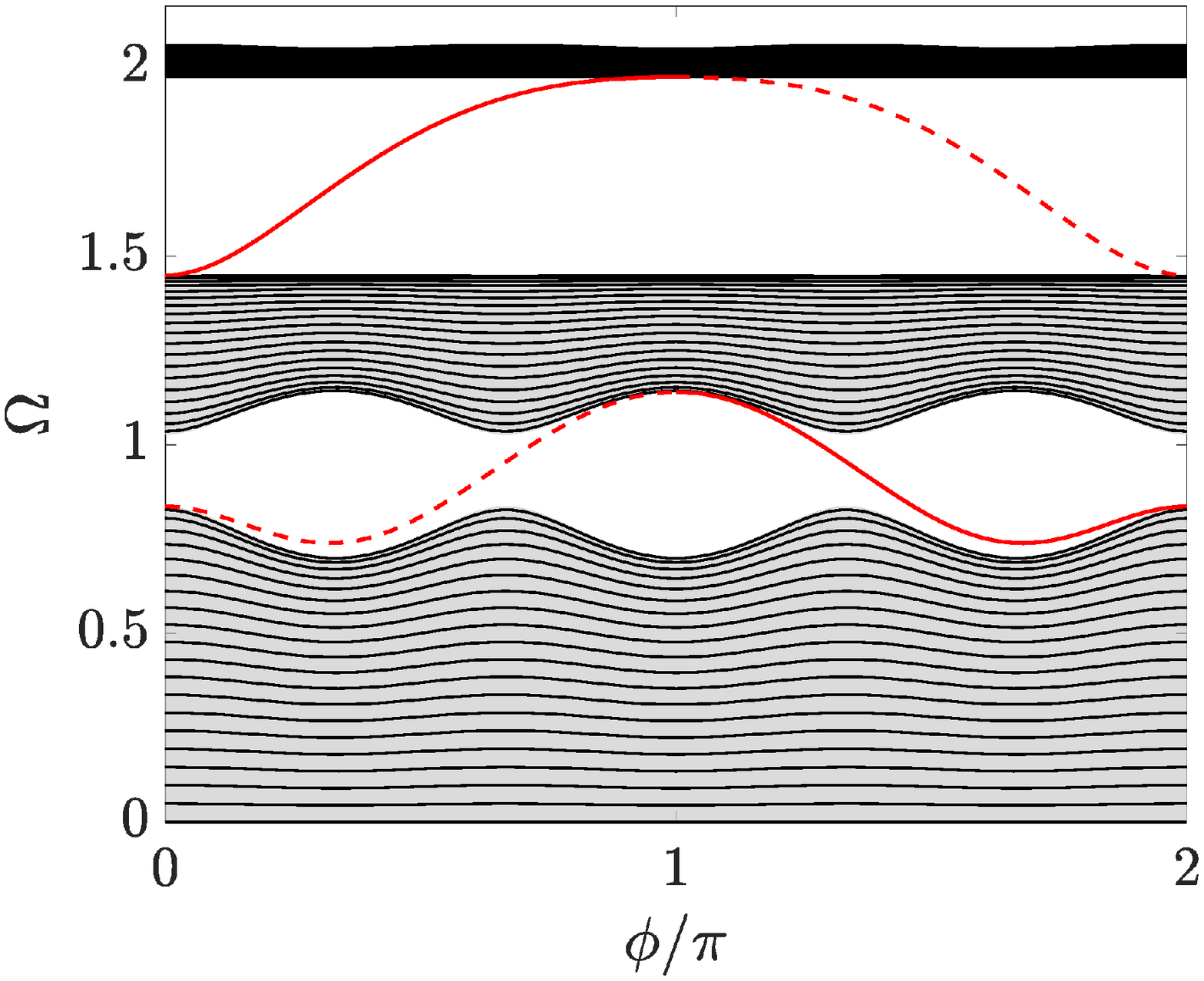}\label{Fig:FiniteChainSpectrum}}
	\subfigure[]{\includegraphics[width=0.45\textwidth]{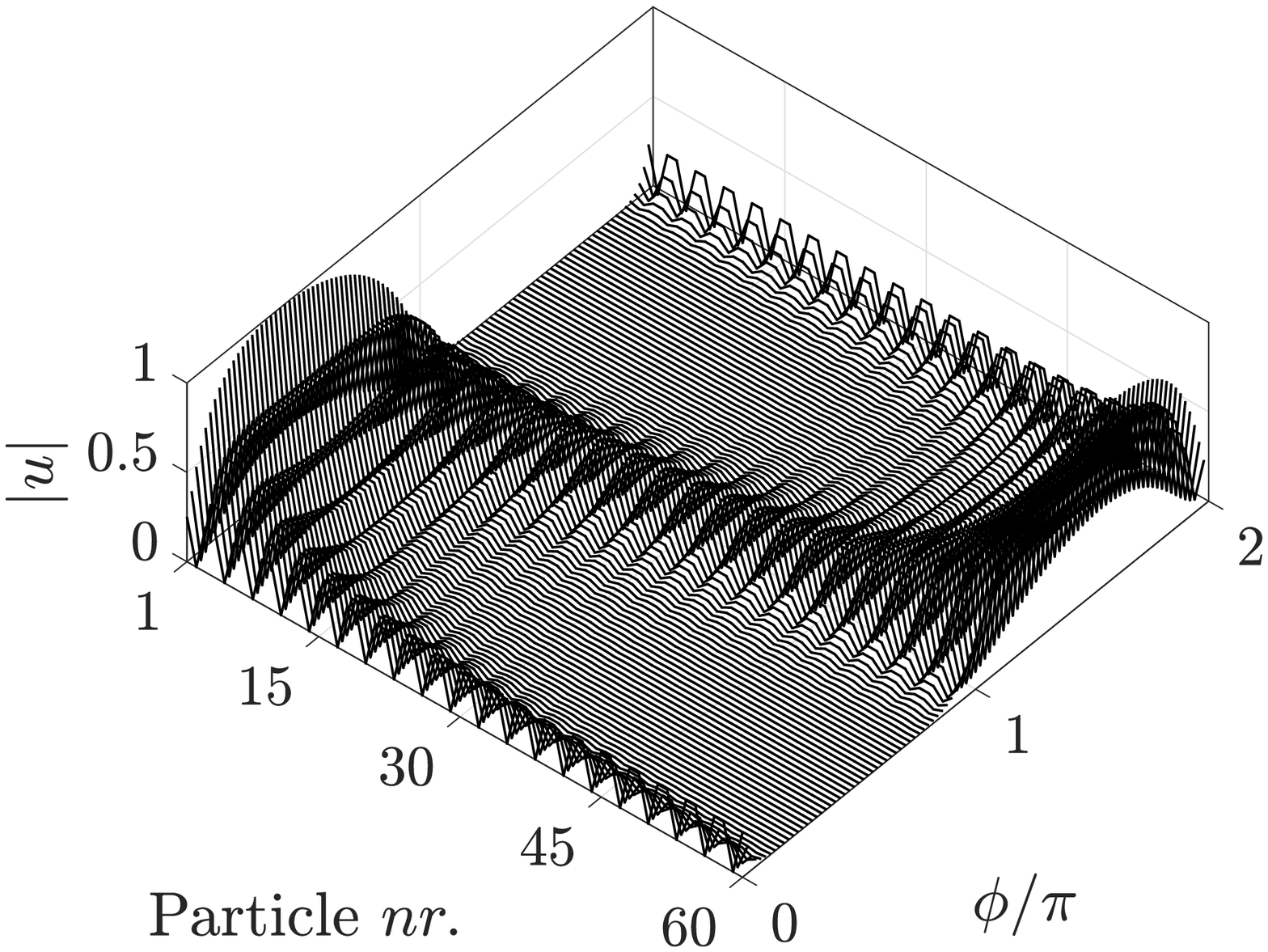}\label{Fig:Chain_edgemode2}}
	\subfigure[]{\includegraphics[width=0.45\textwidth]{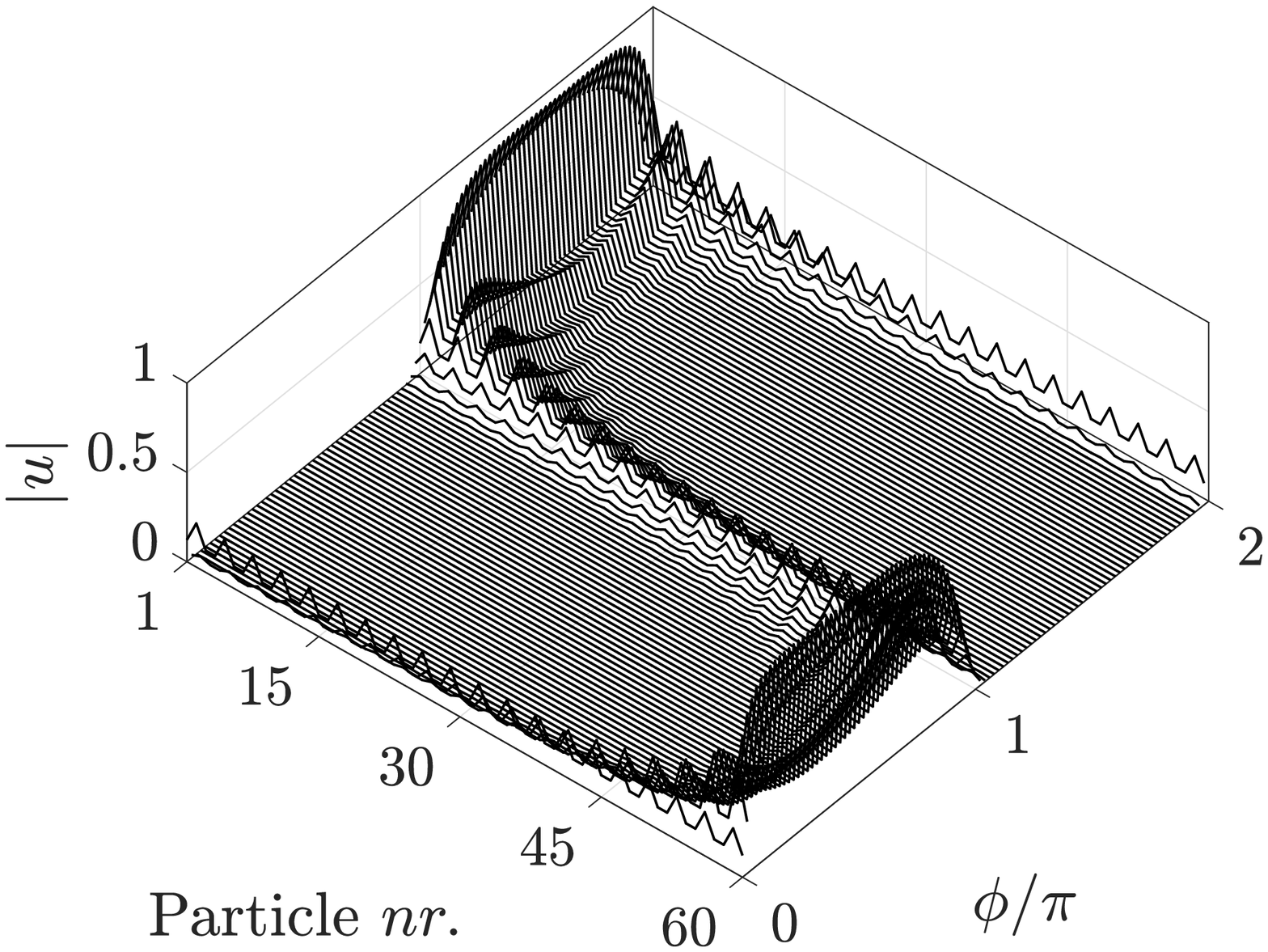}\label{Fig:Chain_edgemode1}}
	\caption{Illustrative example of modes in a ($p/q=1/3$) lattice with modulation parameter $\alpha=0.6$. (a) Dispersion surfaces as a function of $\mu$ and $\phi$ showing the presence of $3$ bulk bands with two bandgaps, along with their corresponding Chern numbers. (b) Natural frequencies of a finite chain of $60$ masses as a function of $\phi$ (black lines) superimposed to the bulk bands (shaded gray regions). Two localized modes (red lines) occupy the two bandgaps, and are represented as dashed lines when localized at the left boundary, and as solid lines when localized at the right. (c,d) Magnitude of the 
		localized mode shapes and their variation with $\phi$ showing how in the first gap the mode transitions from left to right localization, and viceversa for the mode in the second bandgap.}
	\label{Fig:13lattice}
\end{figure}

Here $\bn$ is the outward normal from $\mathcal{D}$ at a point on the boundary $\partial \mathcal{D}$. If $\bA$ were
single valued, then the above integral is zero, since a torus has no boundary. However, since $\bA$ has multiple values 
at $\phi = 0 $ and $2 \pi$, we consider a cut in the torus leading to a cylinder of finite height $2\pi$ along $\phi$. 
Now $\bA$ is single valued in the cylinder and the above integral reduces to evaluating $\bA$ along the two boundaries, $\phi = 0$ and $\phi = 2 \pi$ and for $\mu \in[0,2\pi]$. This gives: 
\begin{equation}
C = \dfrac{1}{2 \pi i} \int_{0}^{2 \pi} \left[ A_{\mu} (\phi = 2 \pi) -  A_{\mu} (\phi = 0) \right] d \mu . 
\end{equation}
with $A_{\mu}  = \bu^* \cdot ( \partial \bu / \partial \mu ) $. Taking the derivative with respect to $\mu$ 
of Eqn.~\eqref{phaseDiff} and its complex conjugate (i.e., ${\bu^*}(\phi=2\pi) = e^{-is\mu}{\bu^*}(\phi=0)$), 
a direct calculation shows that $A_{\mu} (\phi = 2 \pi) -  A_{\mu} (\phi = 0) = i s$. Hence the Chern number results equal to $C = s$. As mentioned before, there are two solutions to the congruence equation for $s$ (Eqn.~\eqref{PSeqn}), corresponding to the two possible stiffness shifts along the lattice. For each band there are two possible values for the Chern number, which is here computed numerically according to the procedure described in~\cite{fukui2005chern}. For the lattice with $p/q=1/3$, this numerical computation gives $C=1$ for the first and third band, and $C=-2$ for the second band (Fig.~\ref{Fig:Disp_unitcell}). From the computed Chern numbers for the individual branches, one can then evaluate and assign labels to each gap, which are defined as the sum of the Chern numbers of all the bands below it~\cite{hatsugai1997topological}. The gap label for gap $r$, here denoted as $C_g^{(r)}$ is therefore given by:
\begin{equation}\label{Eq: gap label}
C_g^{(r)} = \sum_{n=1}^r C_n . 
\end{equation}

The band structure with the corresponding Chern numbers and gap labels for the $p/q=1/3$ lattice is shown in Fig.~\ref{Fig:Disp_unitcell}, where $C_{g}^{(1)}=C_1=1$ and $C_g^{(2)}=C_2+C_1=-1$. 

\subsection{Topological edge modes in a finite chain}\label{EdgeTheorySec}
The Chern numbers and the gap labels are here linked to the presence of topological modes localized at the boundary of finite lattices. To this end, we consider finite lattices with an integer multiple of unit cells and with free boundaries. Such lattice supports topological modes that span the bandgaps as the phase $\phi$ varies, and that are localized at one of the boundaries. These modes are connected to the Chern numbers and gap labels discussed in the previous section. 

Results are presented for a lattice with $N=60$ masses and $p/q=1/3, \alpha=0.6$, which includes $N/q=20$ unit cells. Figure~\ref{Fig:FiniteChainSpectrum} shows the variation of the eigenfrequencies (black lines) of the finite chain as a function of $\phi$. This variation is superimposed to the bulk bands, depicted by the shaded gray areas regions in the figure. The presence of two additional modes (red lines) that span the bandgaps as $\phi$ varies is a notable feature of the finite system spectrum. The solid and dashed lines differentiate modes that are localized at the right and left boundary, respectively. Figures \ref{Fig:Chain_edgemode1} and \ref{Fig:Chain_edgemode2} show the absolute of values $\mathbf{|u|}$ of the spatial variation amplitudes of these modes as a function of $\phi$. We observe that the two modes are localized at one of the two boundaries depending on the value of $\phi$, and that a transition occurs when the corresponding branches touch the bulk bands: the modes become non-localized and transition from being left-localized to right-localized, and viceversa. 

These localized modes occur inside non-trivial bandgaps characterized by nonzero labels and are topologically protected. The bandgap label is given by the sum of the Chern numbers of the bands below the gap. With this definition, the label of a gap measures singularities of the eigenvectors that occur on the top surface of the band below the gap \cite{hatsugai1993edge}, and is connected to the topological modes spanning the gap. Here the absolute value of the gap label $|C_g|$ is related to the number of cycles completed by the associated eigenvalue as it traverses the gap for $\phi \in [0,2\pi]$.  In the $p/q=1/3$ lattice (Fig.~\ref{Fig:13lattice}), both gap labels are $|C_g|=1$, which indicates that the eigenvalues associated with the localized modes in both gaps undergo one cycle for $\phi \in [0,2\pi]$. The sign of the gap label is related to the direction of the transition of the localization during such cycle. A positive label as in the mode in the lower gap corresponds to a transition from right-localized to left-localized (see Fig.~\ref{Fig:13lattice}(c)). The opposite occurs for the mode in the second bandgap, for which a negative label is indicative of a mode transitioning from left to right boundary (see Fig.~\ref{Fig:13lattice}(d)). As mentioned, such transition occurs for the value of $\phi$ for which the branch touches the lower boundary of the gap, i.e. $\phi=0$ and $\phi=2\pi$. Hence, a negative sign of the gap label is related to a mode that transforms from left- to right-localized for increasing $\phi$, with transition occurring when the corresponding branch touches the band below the gap. 

\begin{figure}
	\centering
	\subfigure[]{\includegraphics[width=0.45\textwidth]{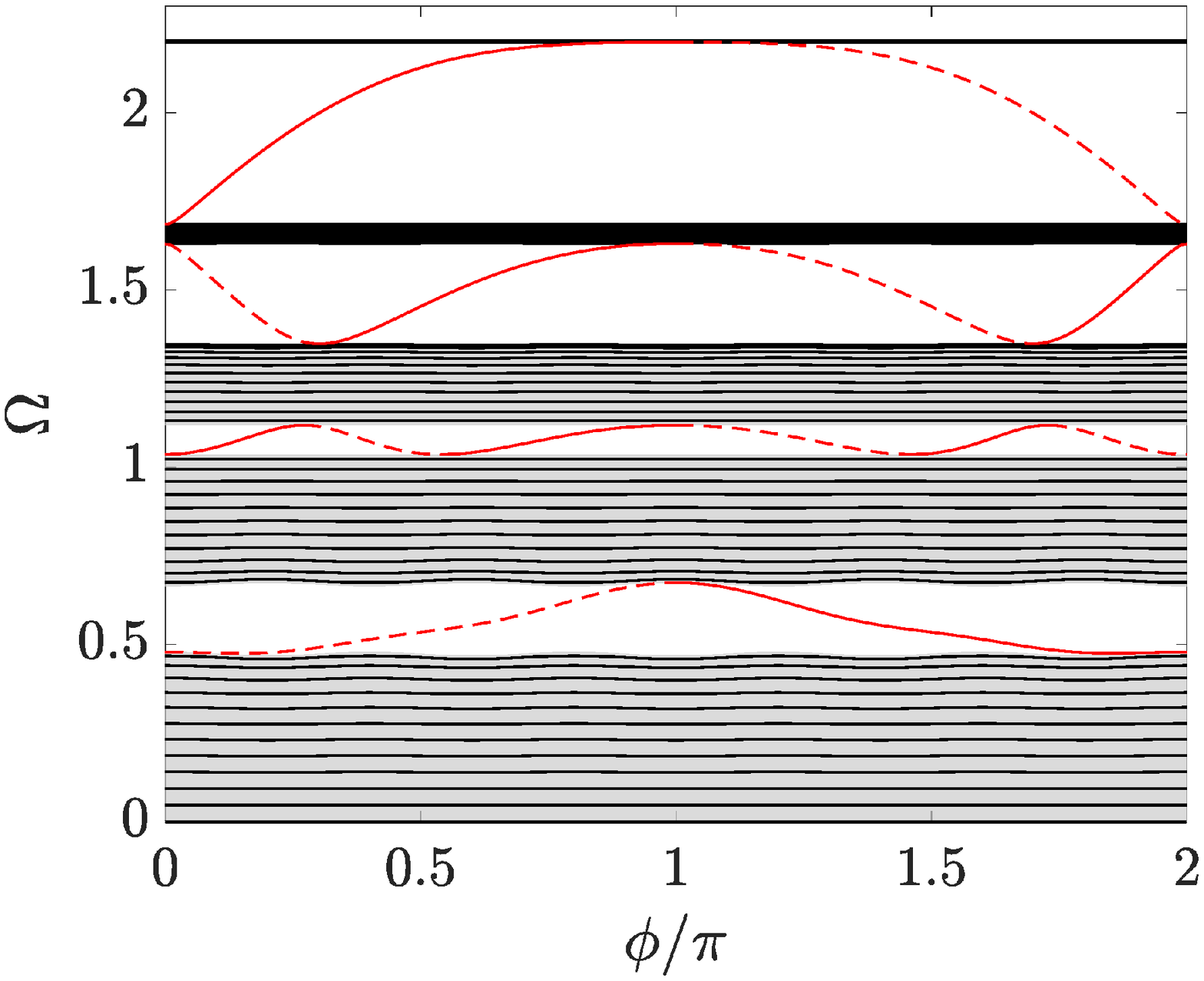}\label{spectrap1q5}}\hspace{0.3cm}
	\subfigure[]{\includegraphics[width=0.45\textwidth]{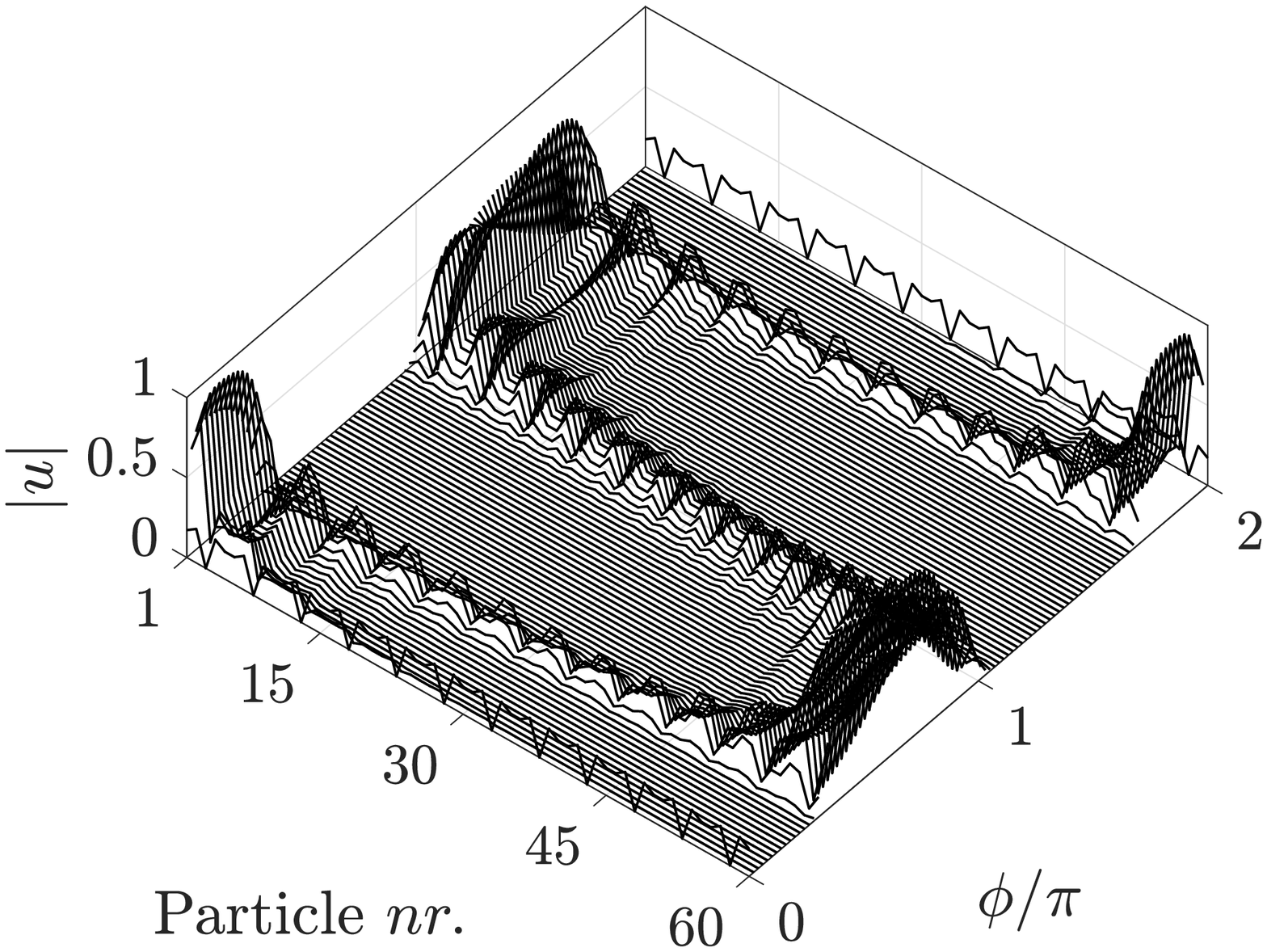}\label{modep1q5}}
	\caption{Illustrative example of modes in a ($p/q=1/5$) lattice with modulation parameter $\alpha=0.6$ and $N=60$ masses. (a) Spectral properties variation as a function of $\phi$ (black lines) superimposed to the bulk bands (shaded gray areas). A single mode is observed to span each gap as a function of $\phi$. Solid and dashed red lines respectively denote modes localized at the right and left boundaries. (b) Magnitude of the topologically protected mode in the third gap and its variation with $\phi$ showing the completion of two cycles as predicted by the gap lable $C_g^{(3)}=-2$.}
\end{figure}

We illustrate another example for the lattice introduced earlier with $p/q=1/5$ and again $N=60$, $\alpha=0.6$. As mentioned, the solution of Eqn.~\eqref{PSeqn} for this case yields $s=1$ or $s=-4$, corresponding to the stiffness shifts of 1 to the left or 4 to the right, respectively. Numerical computation of the Chern number gives $C=1$ for bands 1,3,4 and 5 and $C=-4$ for the second band. The Chern number for the gaps are therefore evaluated as $C_{g}^{(1)}=1$, $C_{g}^{(2)}=-3$, $C_{g}^{(3)}=-2$ and $C_{g}^{(4)}=-1$ (Fig.~\ref{spectrap1q5}), where the gaps are ordered based on increasing frequencies. We note that the gap labels predict the number and nature of the cycles completed by the modes spanning each gap. For example, the mode in the third gap with label $C_g^{(3)}=-2$ complete two cycles as it spans the gap for $\phi \in [0,2\pi]$, as shown in the eigenmode plot of Fig.~\ref{modep1q5}. The mode changes from left-localized to right-localized when it touches the bulk band below the gap with increasing $\phi$, as predicted by the negative sign of its Chern number. 

\section{Topological pumping in 2D lattices}\label{2Dlatticesec}

\begin{figure}[ht!]
\centering
\subfigure[]{\label{fig2Dlattice}
\includegraphics[width=0.4\textwidth]{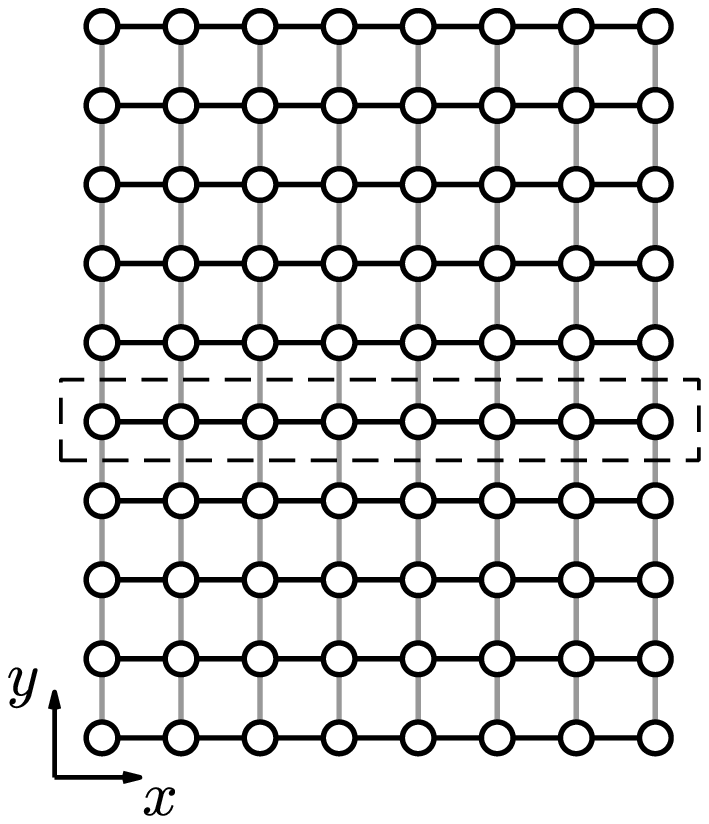}
}
\subfigure[]{\label{figfinitestrip}
\includegraphics[width=0.58\textwidth]{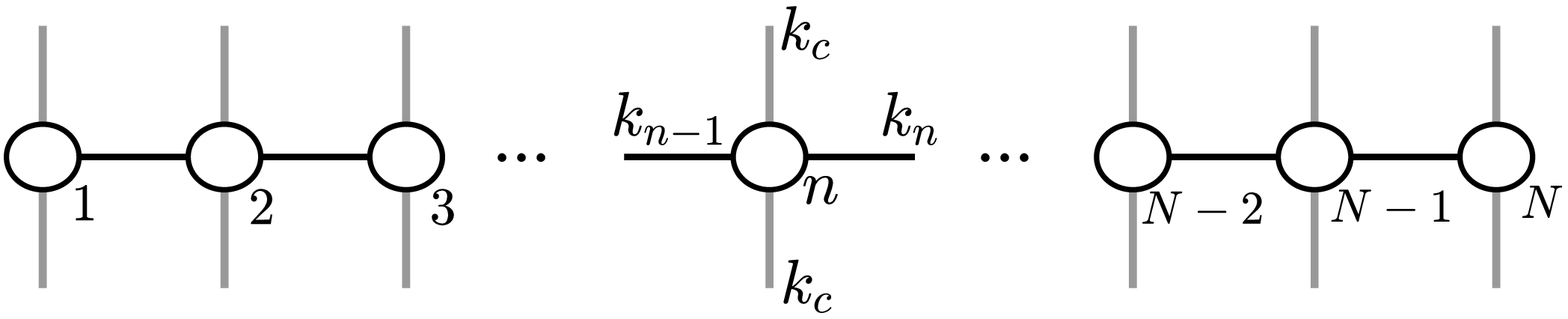}
}
\caption{Schematic of 2D lattice (a), and finite strip with $N$ masses (b). Each spring-mass chain is aligned in $x$ where the horizontal stiffness $k_n$ (represented by black lines) is defined as $k_n=k_0[1+\alpha\cos\left(2\pi pn/q + \phi\right)]$. The chains are coupled along $y$ by springs of constant stiffness $k_c$ (represented by gray lines). }
\end{figure}

We exploit the topologically protected modes and their transition for varying $\phi$ to realize a topological pump. A similar effect was demonstrated for quasi-periodic photonic media in~\cite{kraus2012topological}, and requires the coupling of multiple 1D lattices characterized by smoothly varying $\phi$ along a second dimension. Here, we realize the pumping by stacking the spring-mass chains investigated in the previous section along a second spatial dimension to form a 2D lattice (Fig.~\ref{fig2Dlattice}). In this configuration, the single degree of freedom for each mass now describes its out-of-plane motion, with the springs providing restoring forces that are proportional to the relative transverse motion of neighboring masses. Figure~\ref{figfinitestrip} displays a finite strip of the lattice consisting of a chain with $N$ masses aligned in $x$. The horizontal stiffness $k_n$ connecting the masses (represented by black lines) is defined as previously $k_n=k_0[1+\alpha\cos\left(2\pi p n/q + \phi\right)]$, making the lattice periodic in $x$, with unit cells consisting of $q$ masses, i.e. $k_{n+q}=k_n$. These chains are coupled along the $y$ direction by springs of equal stiffness $k_c$, denoted by the gray lines in Fig.~\ref{fig2Dlattice}. As a result, the modes of the finite 1D lattices can acquire a propagating component along the $y$ direction. Specifically, the topological modes that are localized at one of the boundaries of the 1D chains are transformed into topological wave modes that propagate along $y$, while first being confined to one of the boundaries (left or right in $x$) of the 2D lattice. A topological pump can thus be achieved by gradually, or adiabatically, varying $\phi$ along $y$, that is, by stacking chains with different values of $\phi$ along $y$.  

\subsection{Dispersion analysis of the finite strip}

We investigate the dispersion properties of the finite strip comprised of $N$ masses aligned in $x$ (Fig.~\ref{figfinitestrip}) for assigned $\phi$. We focus on plane wave propagation along $y$ by enforcing Bloch conditions  $u_{n,m+1}=e^{i\mu_y}u_{n,m}$ and $u_{n,m-1}=e^{-i\mu_y}u_{n,m}$, where $n$ and $m$ denote the position of the mass along $x$ and $y$ respectively and $\mu_y$ is the normalized wavenumber of waves traveling along $y$. The governing equation for mass $n,m$ can be written as:
\begin{equation}
	-m\omega^2u_n + [2k_c(1-\cos\mu_y)+k_{n-1}+k_n]u_n - k_{n-1}u_{n-1} - k_nu_{n+1} = 0,
\end{equation}
where the subscript $m$ for the position of the mass along $y$ is dropped for simplicity. Assembly of the $N$ equations for all masses along $x$ yields the eigenvalue problem:
\begin{equation}\label{eq: eig_coupled}
	\mathbf{K}\bm{u}=[\Omega^2-2 \gamma_c(1-\cos\mu_y)]\bm{u},
\end{equation}
where $\bm{u}=\{ u_1, u_2, ..., u_N \}^T$, $\mathbf{K}$ is the $N\times N$ stiffness matrix, $\Omega = \omega/\omega_0$ is the previously defined non-dimensional frequency, while $\gamma_c=k_c/k_0$. Note that $\mathbf{K}$ only depends on the sequence of horizontal stiffnesses $k_n$. Thus the eigenvalue problem in Eqn.~\eqref{eq: eig_coupled} is of the form $\mathbf{K}\bm{u}=\lambda\bm{u}$, and the eigenvalues $\lambda_j$ are those obtained for the finite lattices considered in the previous section for given $\phi$. The corresponding eigenvectors $\bm{u}_j$ define the polarization of wave modes associated with the wavenumber component $\mu_y$. Thus, each eigenvalue $\lambda_j$ defines a dispersion branch which is given by 
\begin{equation}\label{dispfinitestrip}
	\Omega_j(\mu_y)=\sqrt{\lambda_j+2 \gamma_c(1-\cos\mu_y)}.
\end{equation}

\begin{figure}[ht!]
	\centering
	\subfigure[]{\label{dispkc05}
		\includegraphics[height=0.4\textwidth]{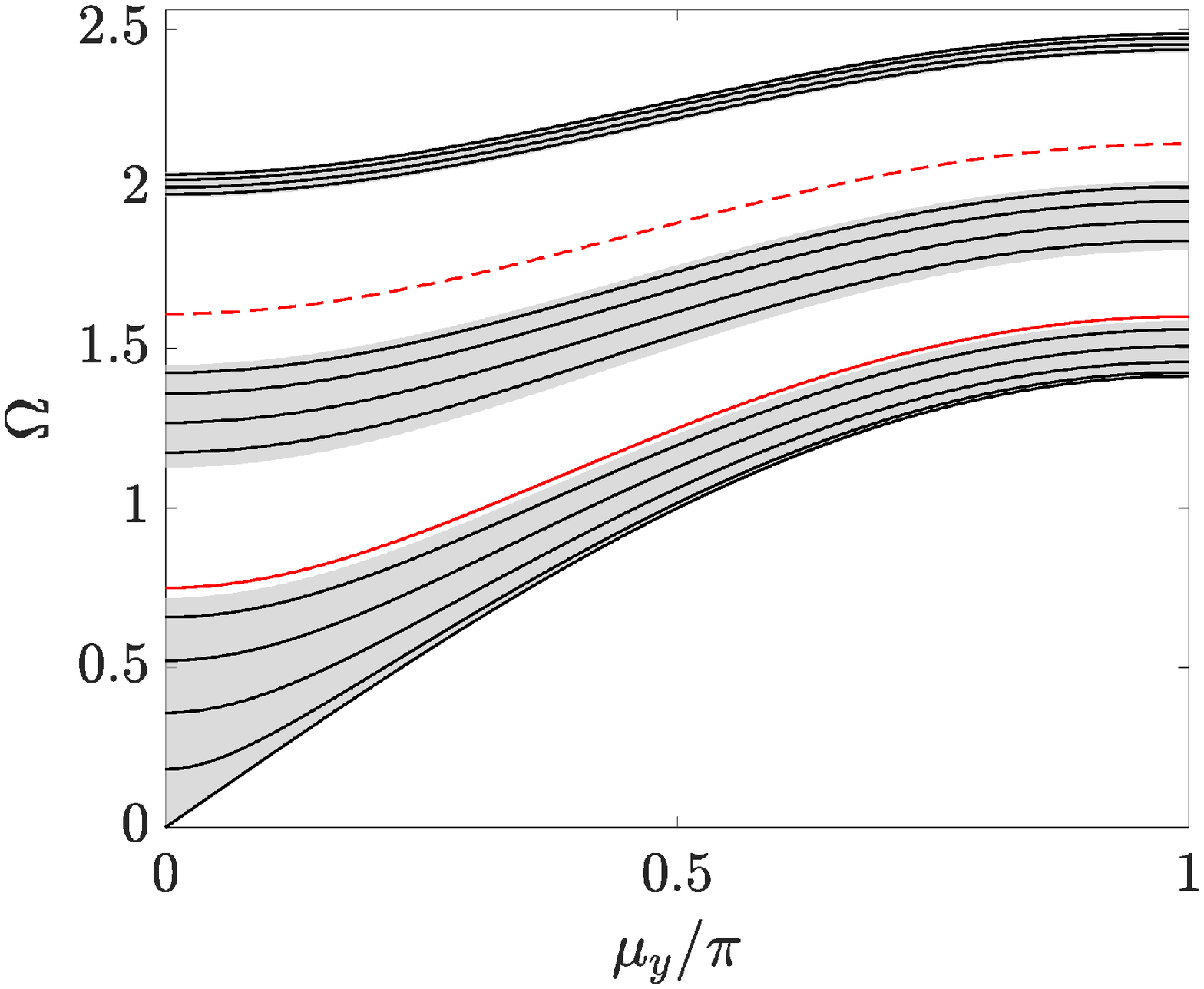}
	}
	\subfigure[]{\label{dispkc2}
		\includegraphics[height=0.4\textwidth]{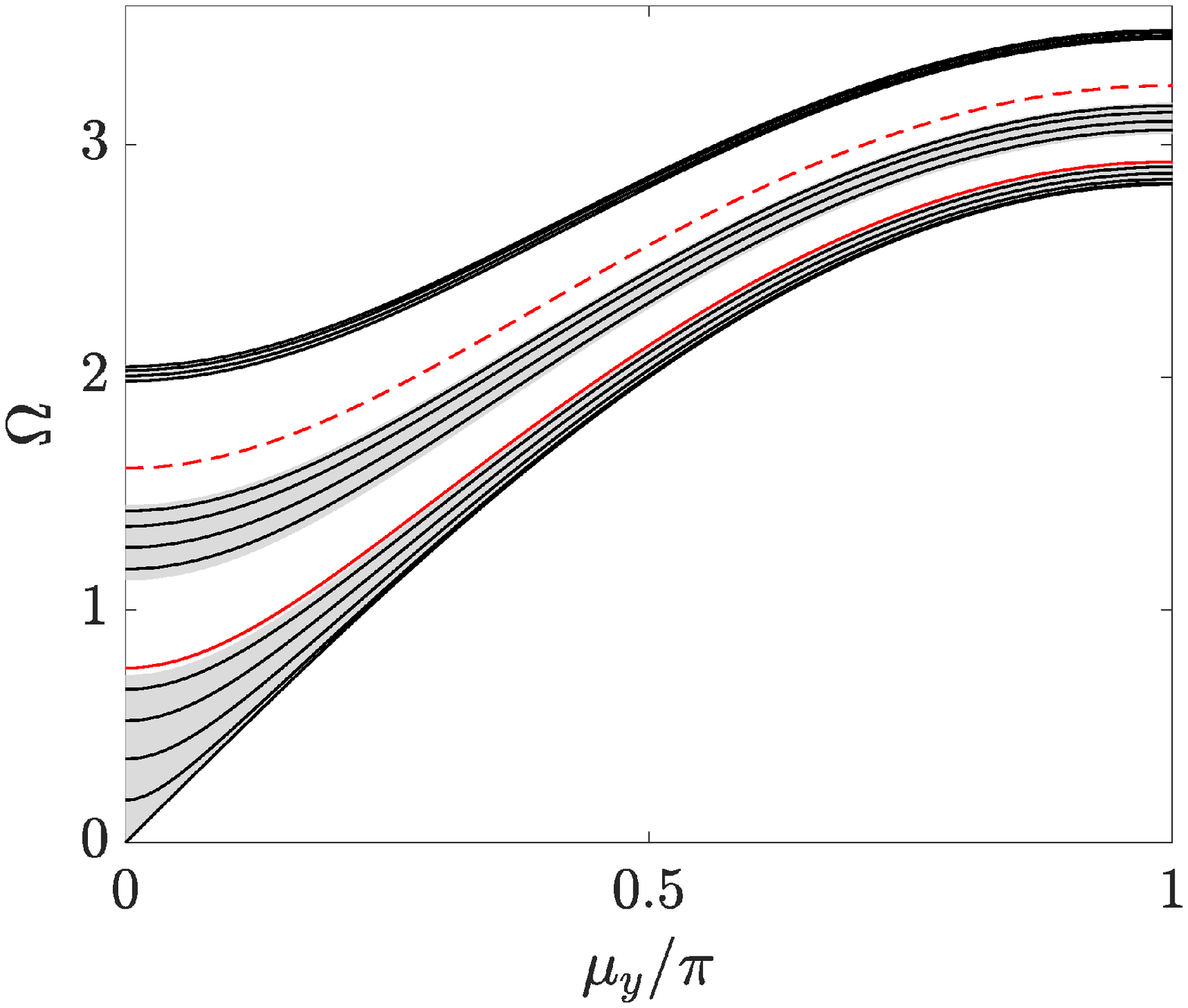}
	}
	\caption{Dispersion properties of finite lattices with $p/q=1/3$, $\alpha=0.6$, $N=15$ masses and $\phi=1.75\pi$:  $\gamma_c=0.5$ (a), and $\gamma_c=2$ (b). The finite lattice bulk modes represented by black lines are superimposed to the bulk spectra (shaded gray regions). Wave modes localized at the left and right boundaries are represented by dashed and solid red lines, respectively.}
	\label{dispersionfinitestrip}
\end{figure}

Figures~\ref{dispkc05} and \ref{dispkc2} show the dispersion relations for two lattices respectively with $\gamma_c=0.5$ and $\gamma_c=2$ and $p/q=1/3$, $\alpha=0.6$, and $\phi=1.75\pi$. The bulk properties are first evaluated by considering lattices of infinite extent in both $x$ and $y$, which identifies the three bulk regions represented by the shaded gray areas in the figures. Wave modes for strips including $N=15$ masses along $x$ are superimposed to the bulk spectra as solid black lines, while the red lines inside the gaps identify edge modes. Again, modes localized at the left or right boundary are respectively represented by dashed and solid lines.

\begin{figure}
	\centering
	\subfigure[]{\label{omegamuzero}
		\includegraphics[width=0.4\textwidth]{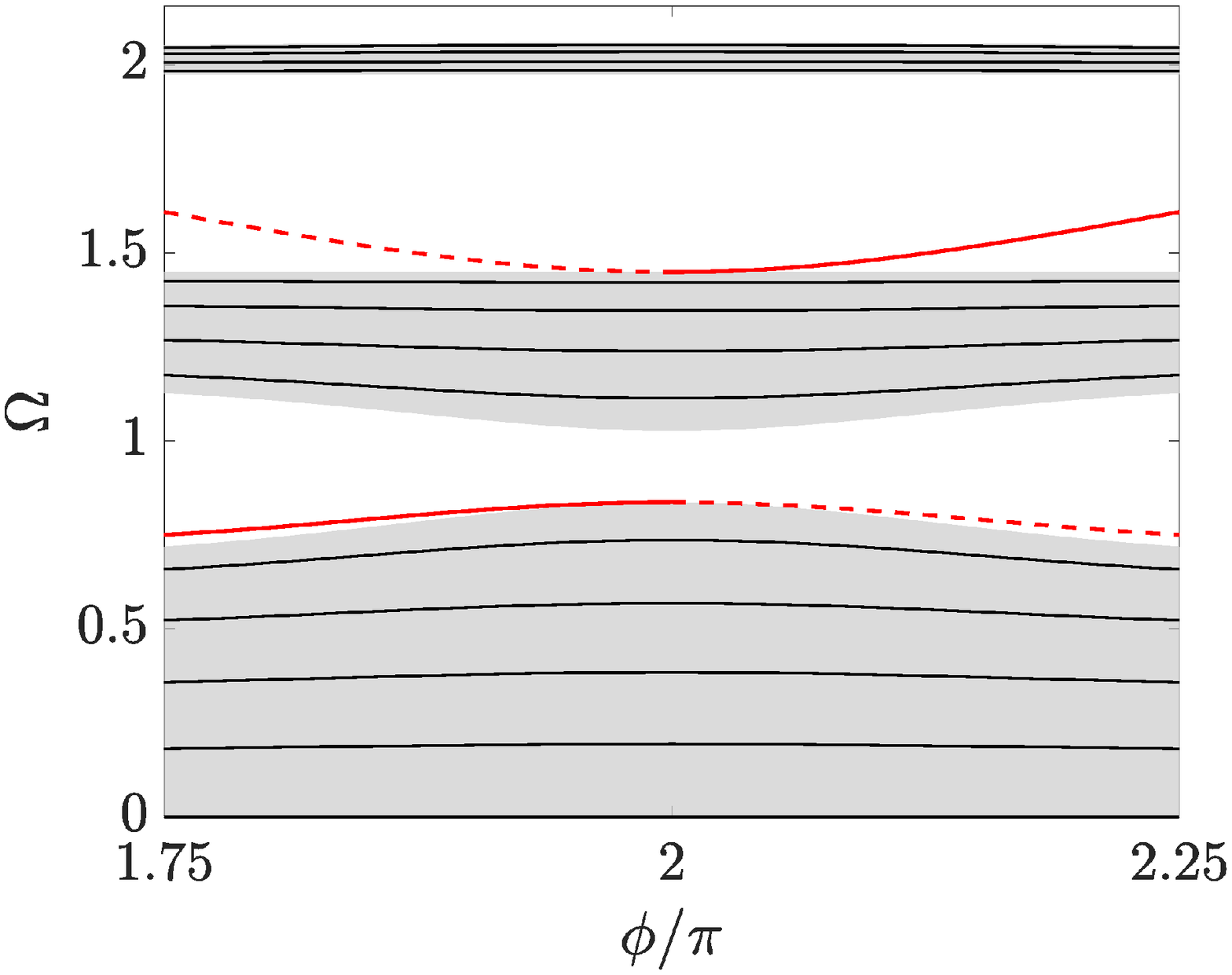}
	}
	\subfigure[]{\label{omegamuhalf}
		\includegraphics[width=0.4\textwidth]{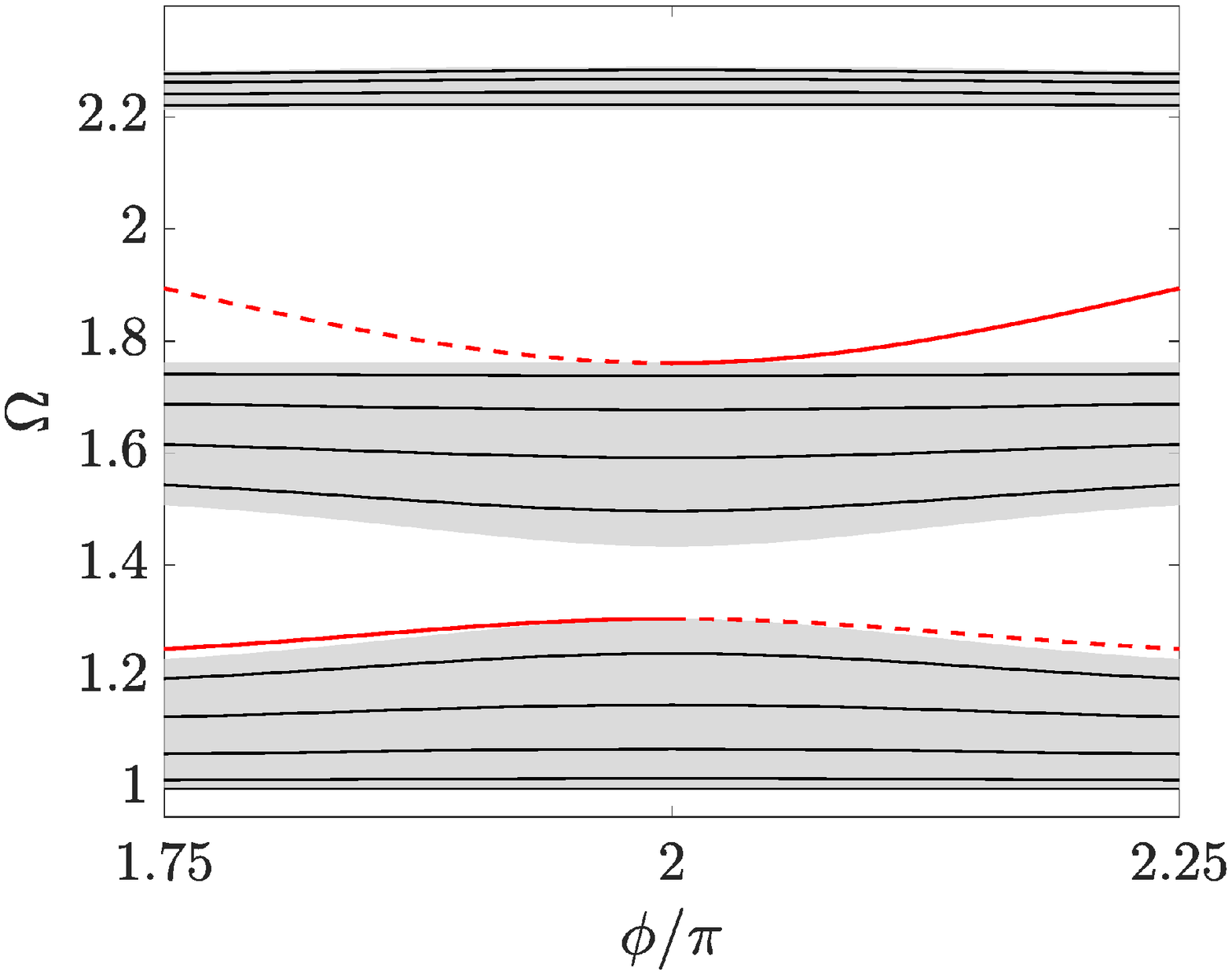}
	}
	\subfigure[]{\label{modesphi}
		\includegraphics[width=0.4\textwidth]{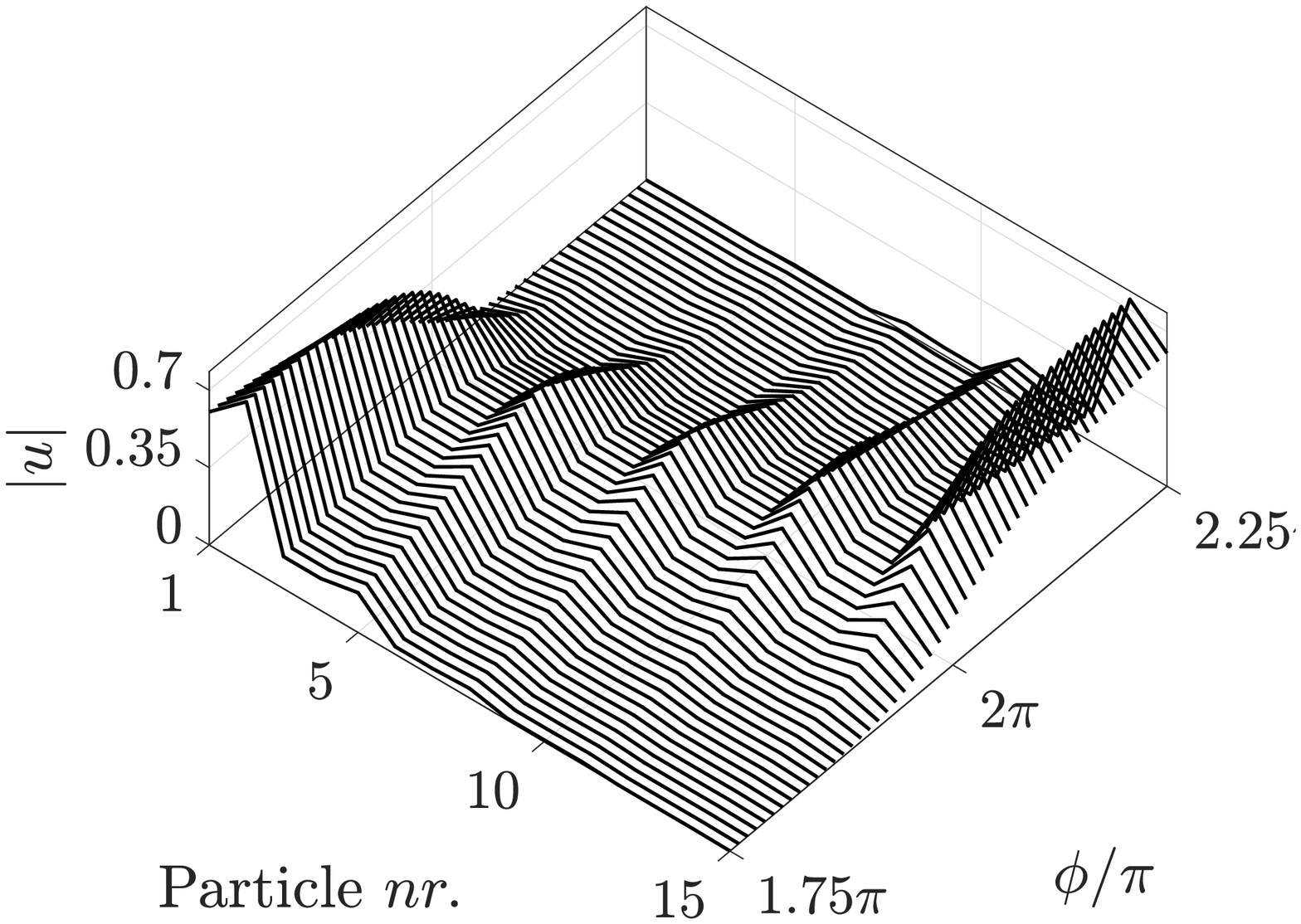}
	}
	\caption{Dispersion branches variation in terms of $\phi$ for $\mu_y=0$ (a), and $\mu_y=0.5\pi$ (b) for the finite strip with $p/q=1/3$, $\alpha=0.6$, $N=15$ masses and $\gamma_c=0.5$. The finite lattice bulk modes are represented by the black lines and superimposed to the bulk spectra (shaded gray regions), while the localized modes spanning the gaps are represented by red lines. Magnitude of the second gap topological eigenmode as a function of $\phi$ showing the left-to-right transition.}
\end{figure}

Next, we show the dispersion branches for $\gamma_c=0.5$ for $\phi\in [1.75\pi, \,\, 2.25\pi]$. This choice is motivated by the topological pumping process described in the next section, with the observation that the interval $\phi \in [2\pi,2.25\pi]$ is the same as $\phi \in [0,0.25\pi]$ due to the periodicity of $k_n$ with $\phi$. Figures~\ref{omegamuzero} and \ref{omegamuhalf} follow the evolution of the dispersion of the branches with $\phi$ for  $\mu_y=0$ and $\mu_y=0.5\pi$, respectively. Similarly to what was illustrated for decoupled 1D lattices ($\gamma_c = 0$), even in the presence of coupling, i.e. for $\gamma_c\neq0$, variations in $\phi$ cause the localized modes to transition from one edge to the other. This transition occurs as waves propagate along $y$ according to the specific value of $\mu_y$. The evolution of the eigenmode in the second gap as a function of $\phi$ is illustrated in Figure~\ref{modesphi}, where a transition from left-localized to right-localized occurs at $\phi=2\pi$. This mode and its transition is employed next to realize the pump described in the next section.

\subsection{Forced response and topological pumping}
The topological pump consists in waves undergoing a left-to-right edge transition through the excitation of edge modes and the modulation of the stiffness constants along the propagation direction. To this end, we consider a 2D lattice of $M$ coupled 1D chains, each characterized by specified values of the stiffness phase modulation $\phi$. The topological pump is sought to be produced by gradually, or adiabatically, modulating $\phi$ along $y$ which causes a mode initially localized at one of the boundaries to evolve its $x$ polarization as it travels along $y$. This is expected to occur under the assumptions stated by the adiabatic theorem, which essentially implies that the stiffness is modulated gradually enough to allow the wave modes to adapt to the local properties of the lattice~\cite{nassar2018quantization}. This process is akin to the adiabatic evolution of an eigenstate in quantum mechanics, where here the propagation direction $y$ takes the role of time according to a similar effect previously demonstrated in photonic quasicrystals~\cite{kraus2012topological}. 

The pumping process begins by exciting a selected topological mode of the system, which is done by imposing frequency and a polarization in $x$. Specifically, a left-localized mode polarization identifies the second gap topological mode, while the choice of $\Omega$ defines the associated wavenumber $\mu_y$ and corresponding group velocity components along $y$. As the resulting waves propagate along $y$, their polarization along $x$ adapts to the mode shape of the topological mode at the local value of $\phi(y)$. This is enabled by the slow variation of the modulation: a high rate of change of $\phi$ with $y$ would in contrast prevent the distribution of displacements along $x$ from changing significantly over a short distance in $y$, and would cause conversion to other wave modes with branches at the considered frequency. Hence, waves traveling faster along $y$ require lower rates of change of $\phi(y)$ to allow for the desired adiabatic evolution. 

\begin{figure}[ht!]
	\centering
    \includegraphics[height=0.6\textwidth]{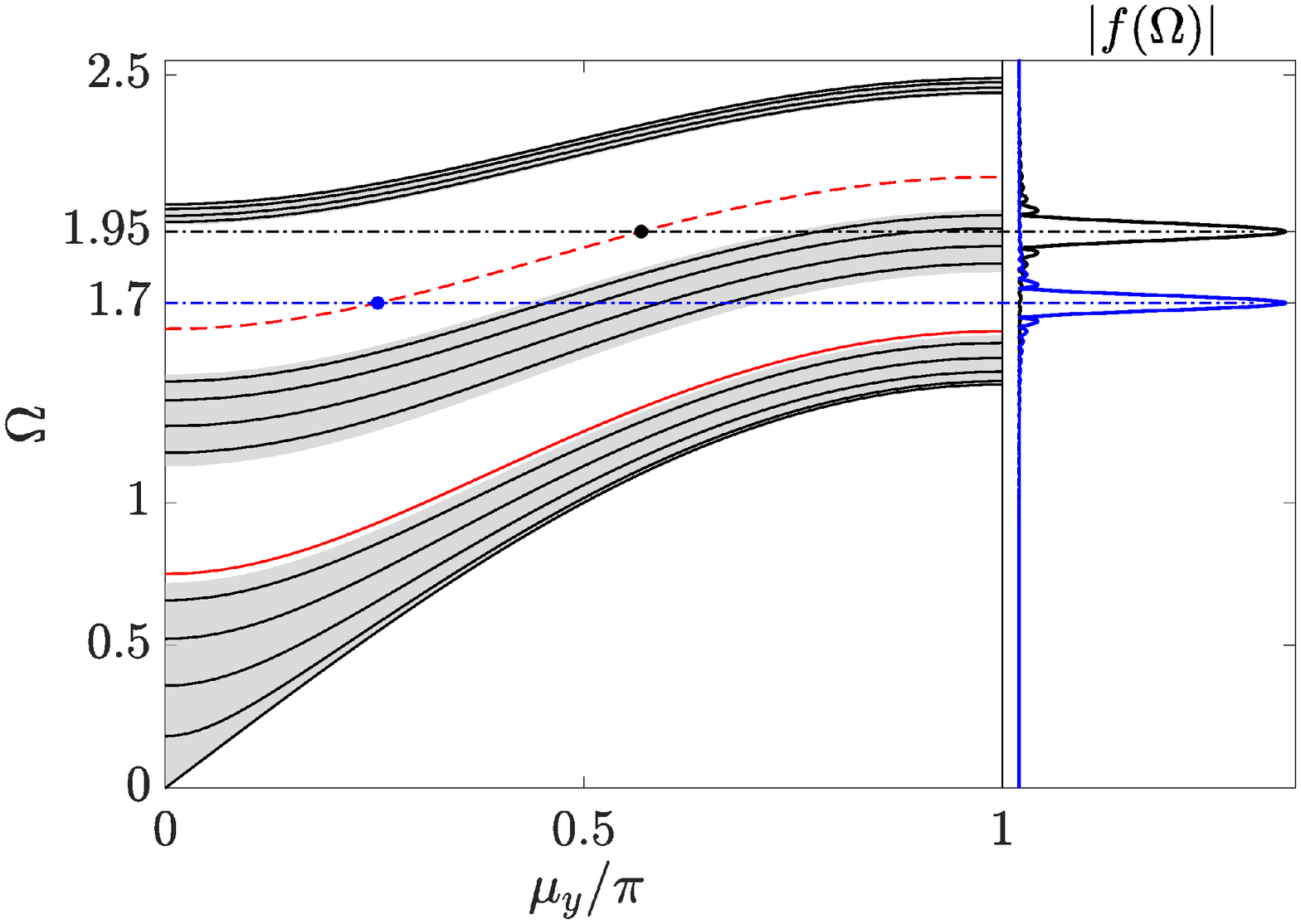}
	\caption{Excitation spectra at $\Omega=1.7$ and $\Omega=1.95$ for excitation of the left-localized topological edge mode. Here, $|f(\Omega)|$ denotes the magnitude of the Fourier Transform of the excitation.} 
    \label{dispersionexcitation}
\end{figure}

We demonstrate this concept by computing the transient response of a finite 2D lattice excited by band limited sinusoidal pulses at selected frequencies. The 2D lattice is formed by stacking $M=150$ chains along $y$ of the kind considered before ($p/q=1/3$, $\alpha=0.6$, $N=15$ masses, and $\gamma_c=0.5$). The choice of $\gamma_c$ between the two cases considered in Fig.~\ref{dispersionfinitestrip} is motivated by the lower group velocity along $y$ that can be inferred from the slope of the dispersion branch of the topological modes. Accordingly, the distance required for the transition is expected to be shorter. Also, we target the localized mode in the second gap, which is wider and features a topological mode with stronger localization in comparison to the first gap (see Figs.~\ref{Fig:Chain_edgemode1} and \ref{Fig:Chain_edgemode2}). Here $\phi$ is varied from $\phi_1=1.75\pi$, where the mode is localized at the left edge, to $\phi_2=2.25\pi$, where the mode is localized at the right edge. Two sine burst signals (amplitude modulated harmonic signals) with carrier frequencies $\Omega=1.7$ and $\Omega=1.95$, both with sinusoidal amplitude modulation spanning 50 cycles, are considered. The spectra of these narrowband signals are displayed in Fig.~\ref{dispersionexcitation} alongside the dispersion branches of the strip with $\gamma_c=0.5$. We note that excitation of bulk modes existing at these frequencies is minimized through the enforcement of the $x$-wise mode polarization corresponding to the topological mode of interest, which is imposed by prescribing the motion of lattice at $y=1$.

\begin{figure}
\centering
\subfigure[]{\label{phiconstant}
\includegraphics[height=0.4\textwidth]{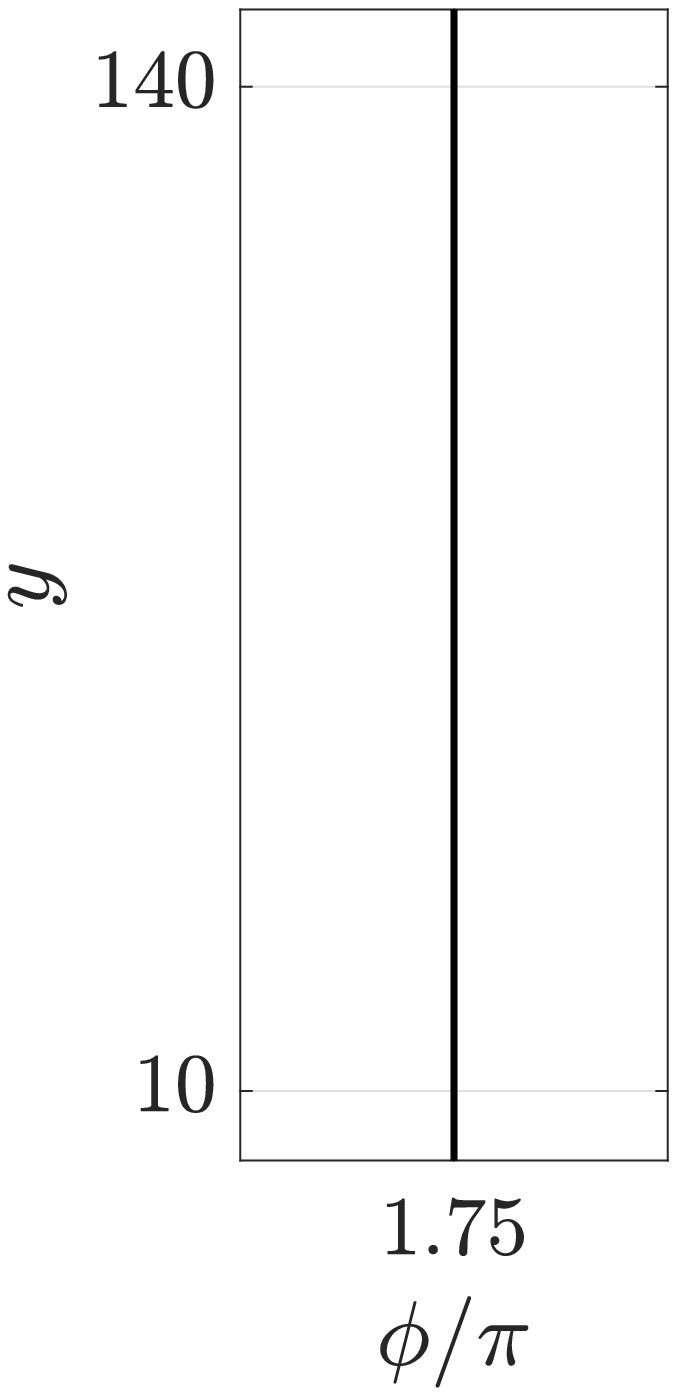}}
\subfigure[]{\label{w17phiconstant}
\includegraphics[height=0.4\textwidth]{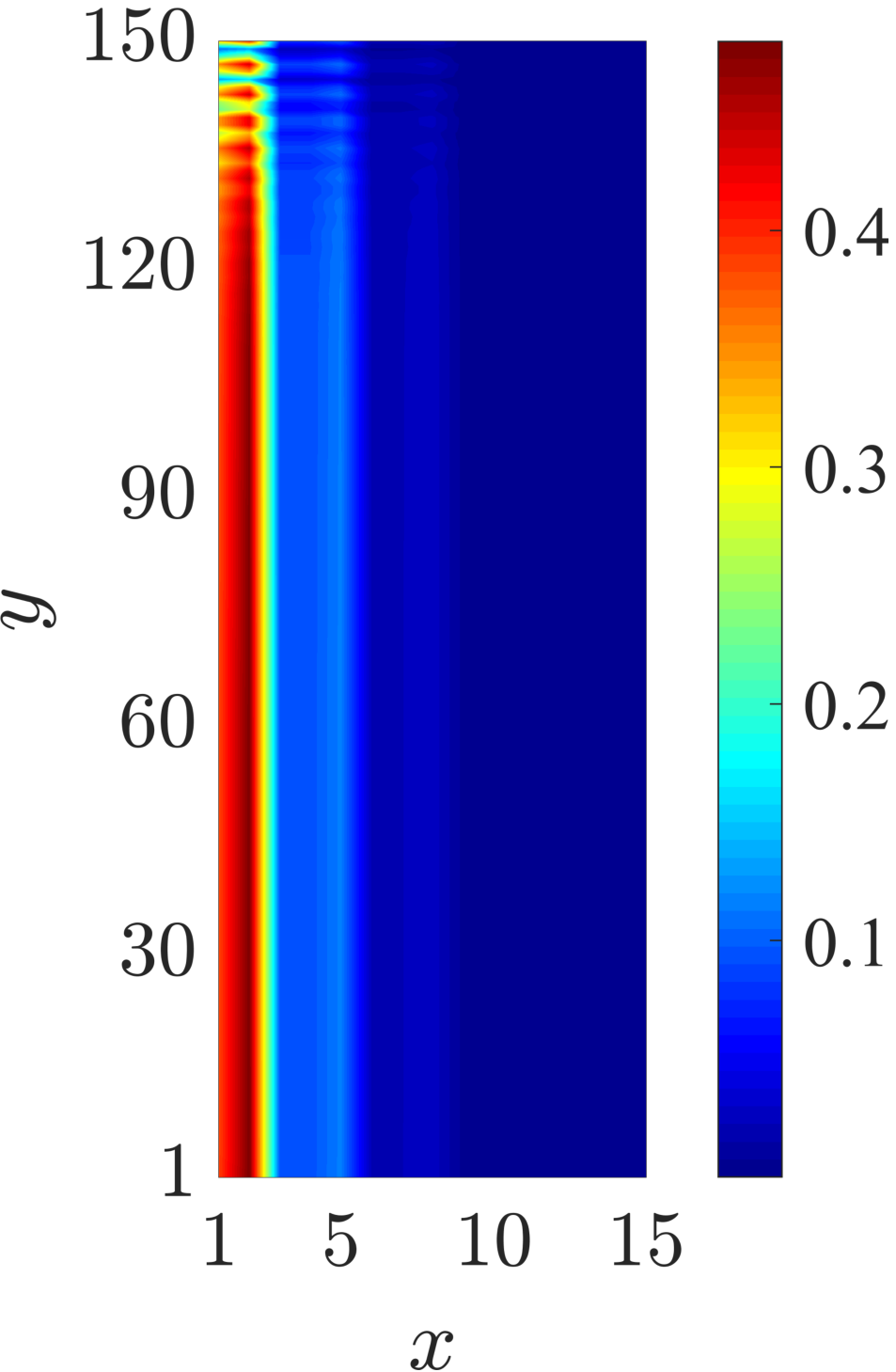}}
\subfigure[]{\label{w195phiconstant}
\includegraphics[height=0.4\textwidth]{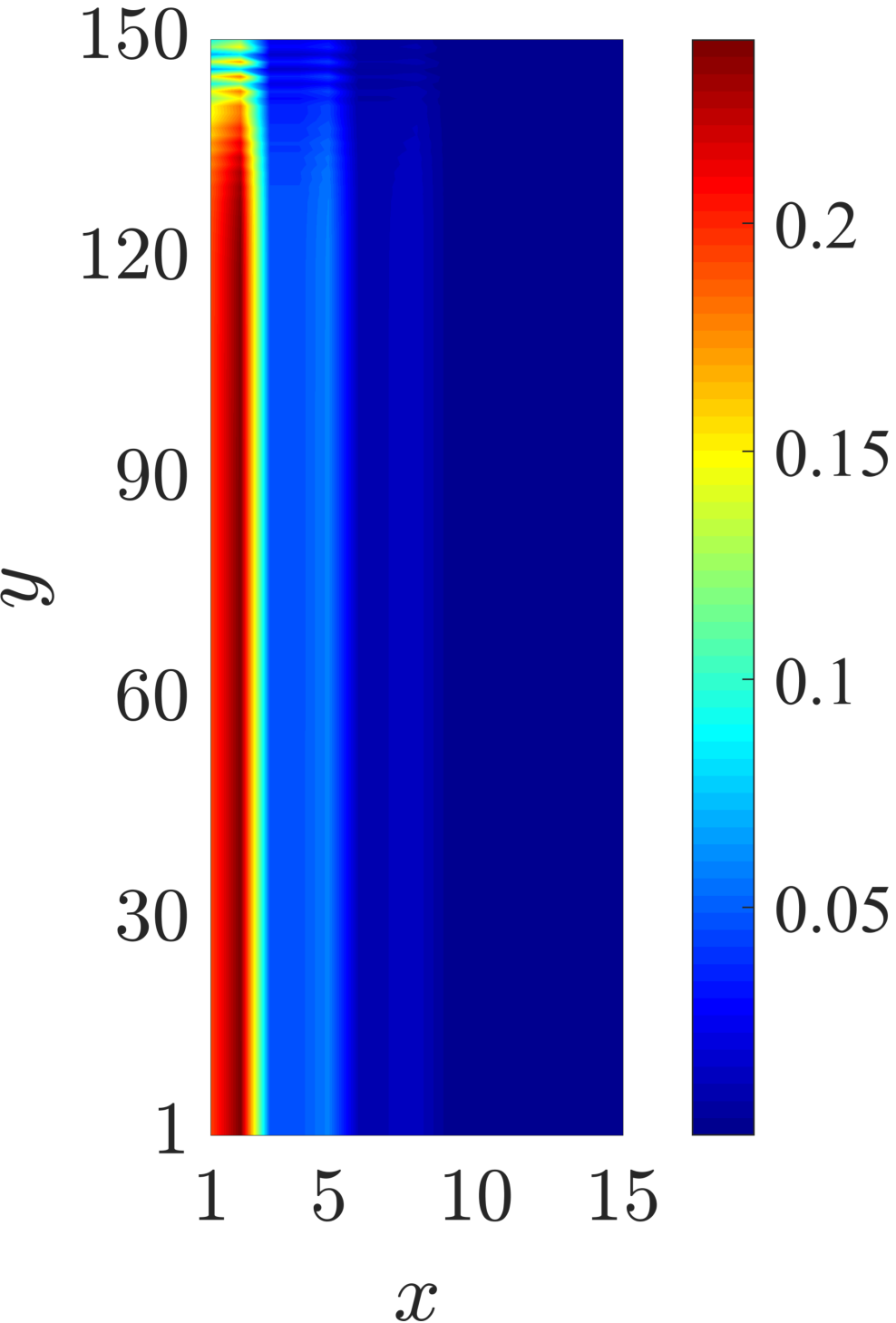}}\\
\subfigure[]{\label{phivaries}
\includegraphics[height=0.4\textwidth]{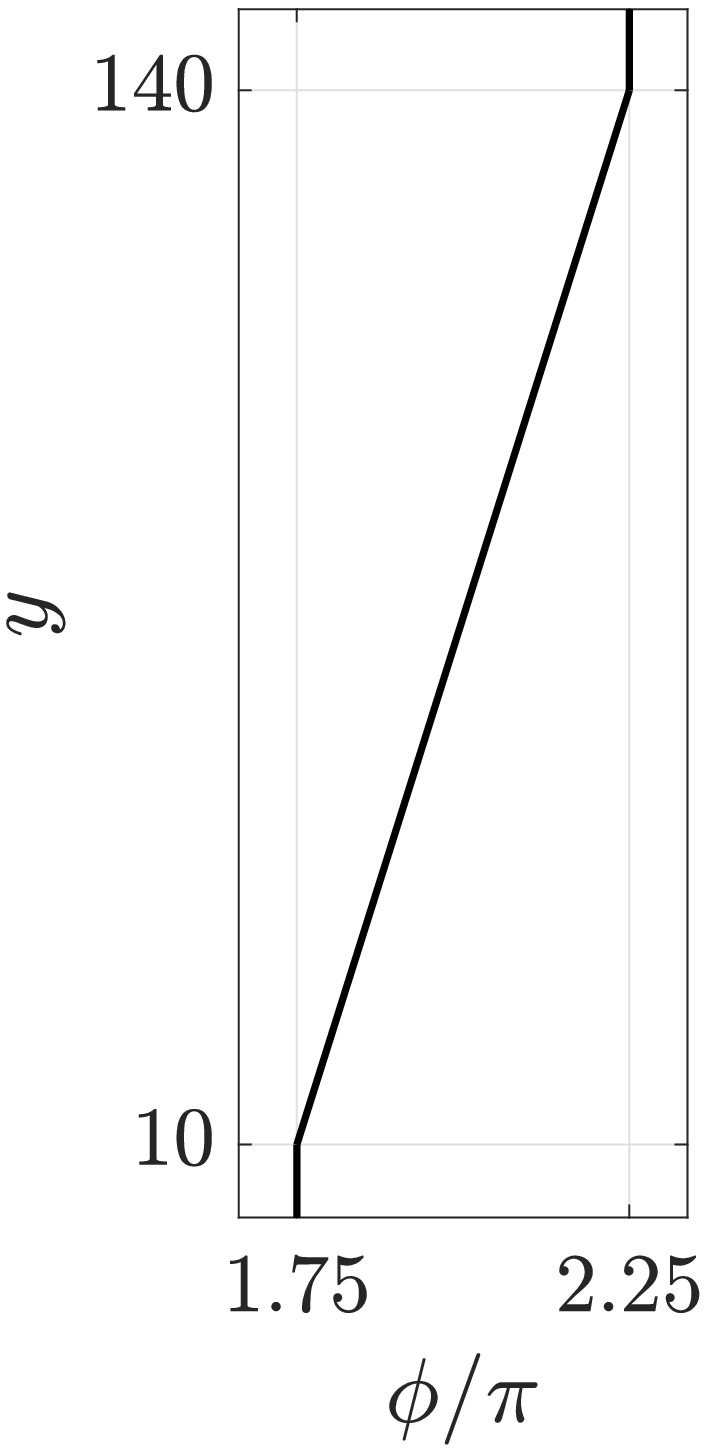}}
\subfigure[]{\label{w17phivaries}
\includegraphics[height=0.4\textwidth]{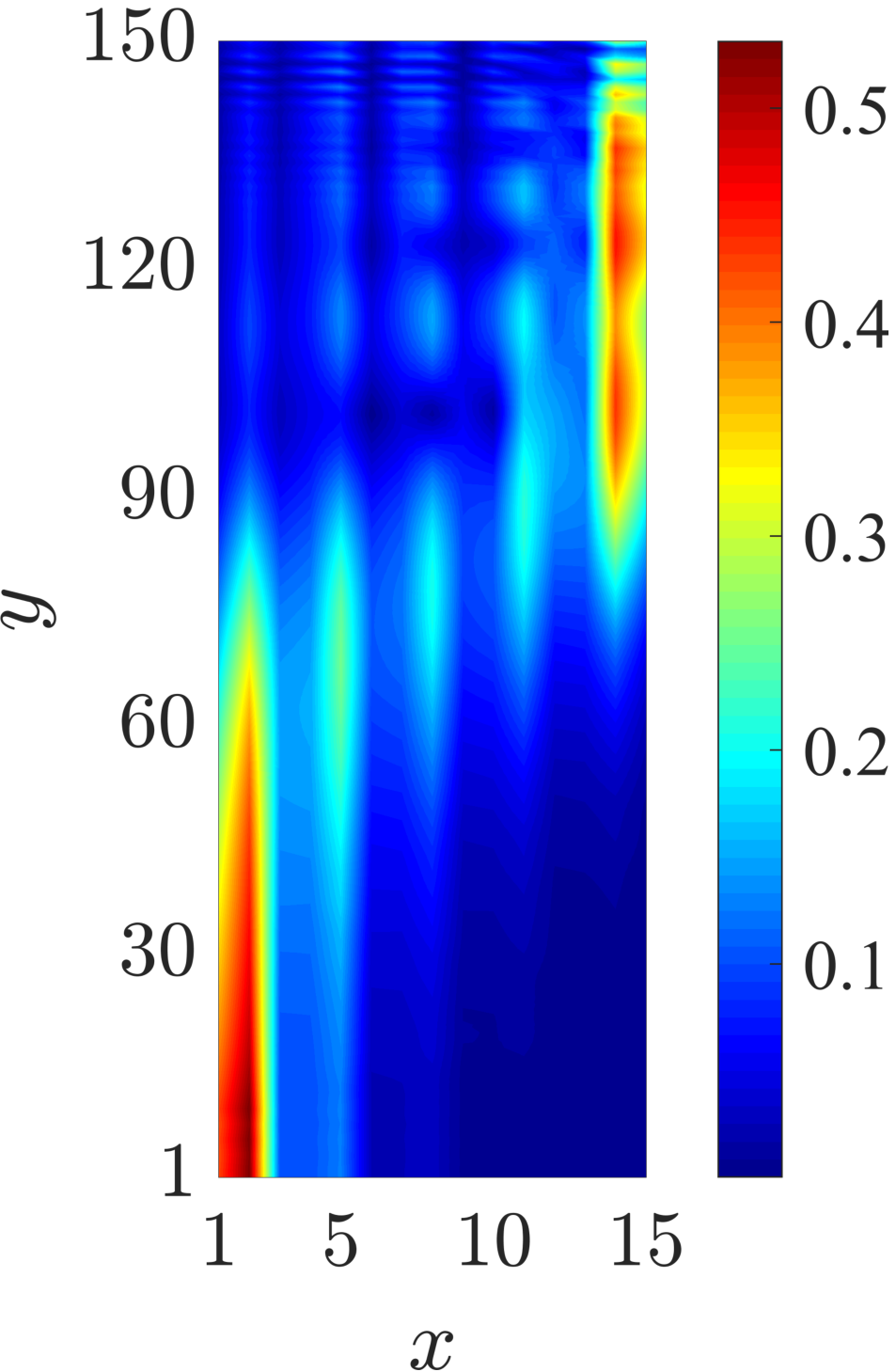}}
\subfigure[]{\label{w195phivaries}
\includegraphics[height=0.4\textwidth]{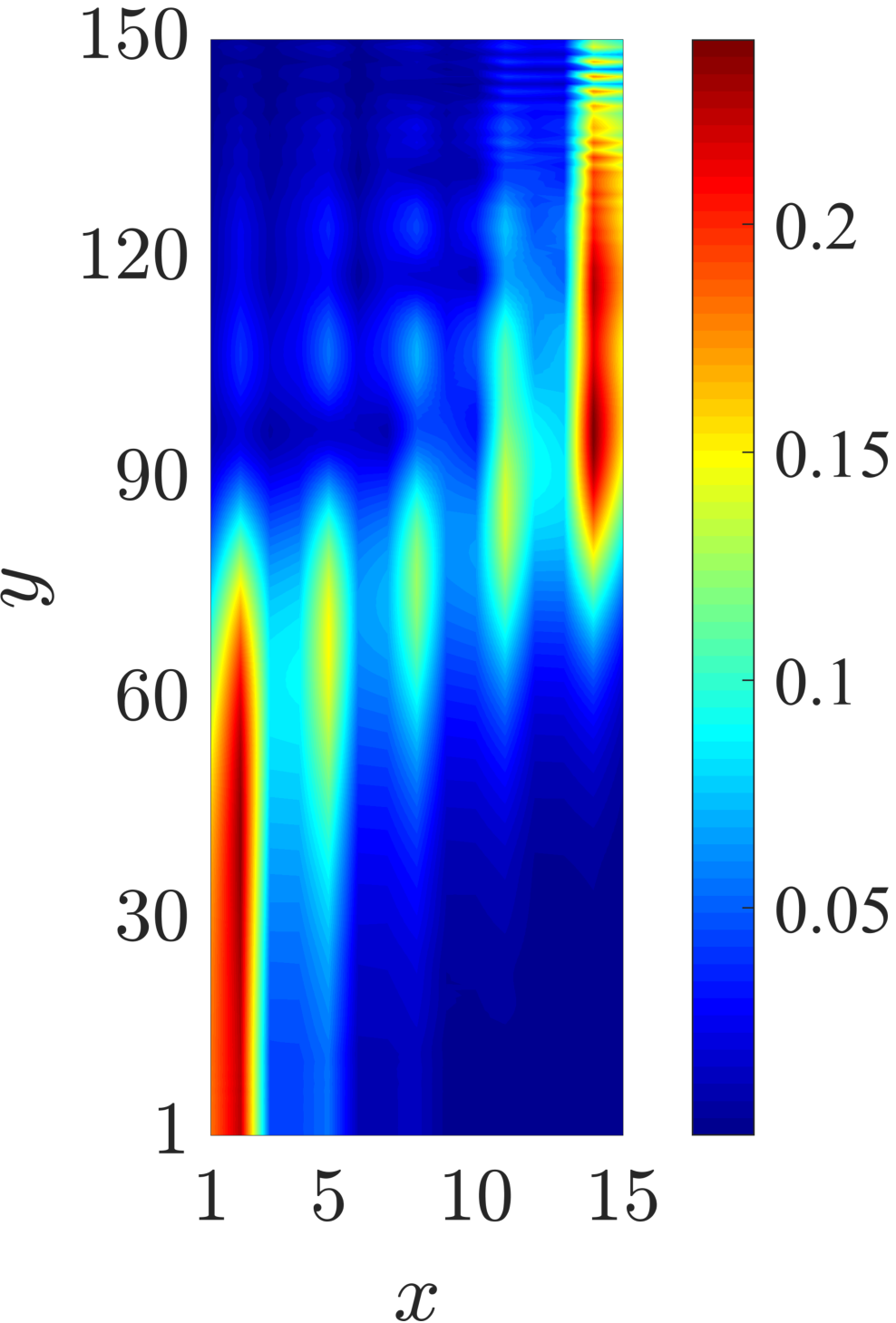}}
\caption{Transient response of 2D lattice to band-limited sinusoidal pulses. (a) Phase of modulation $\phi(y)$ for results shown in (b,c) where $\phi$ is constant to $\phi_1=1.75\pi$ on the entire lattice. (b,c) RMS of displacement field for excitation at $\Omega=1.7$ and $\Omega=1.95$ respectively. (a) Phase of modulation $\phi(y)$ for results shown in (e,f) where $\phi$ is varied along $y$ to allow for the adiabatic evolution causing the pump. (e,f) $||\bm u||_2(x,y)$ field for excitation at $\Omega=1.7$ and $\Omega=1.95$ respectively.}
\label{transientpump}
\end{figure}

The simulation results are summarized in Fig.~\ref{transientpump}. The transient response of the lattice was simulated for both excitation frequencies ($\Omega=1.7$ and $\Omega=1.95$) and for both $\phi$ constant and $\phi$ varying linearly along the middle portion of the $y$ dimension (Fig.~\ref{transientpump}.(a,d)). The top figures (b,c) correspond to the case of  $\phi$ kept constant at $\phi=1.75\pi$ (Fig.~\ref{phiconstant}), while the bottom figures (e,f) are obtained for $\phi$ varying linearly from $1.75\pi$ to $2.25\pi$ in the range $y \in [11,140]$. The onset of the desired mode is facilitated by keeping $\phi$ constant for the beginning and concluding portions of the lattice (10 chains) as illustrated in Fig.~\ref{phivaries}. The colormaps shown in figures (b,e) (middle) show the $\mathcal{L}^2$ norm in time $||\bm u||_2(x,y)$ of the response for  excitation at $\Omega=1.7$ while figures (c,f) (right) correspond to $\Omega=1.95$. We observe that for constant $\phi$ (Figs.~\ref{transientpump}(b,c)) the waves travel along $y$ while remaining confined to the left boundary due to the excitation of the left-localized edge mode. In contrast, pumping occurs when $\phi$ varies smoothly along $y$ (Figs.~\ref{transientpump}(e,f)) causing the transition from left-localized to right-localized wave propagation due to the adiabatic evolution of the edge wave.

\section{Conclusions}\label{ConcSec}

The topological properties of 1D lattices with periodic stiffness modulation are investigated to implement a topological pump for elastic waves in 2D discrete elastic lattices. A family of 1D elastic lattices with periodic stiffness modulation are characterized by non-trivial gaps that are spanned by topological modes localized at the boundaries of finite lattices. Coupling and stacking of 1D lattices along a second dimension allows the adiabatic modulation of the wave properties, which enables the transition of topologically protected wave modes from one edge to the other of the domain. The concept presented herein describes the conditions which characterize the existence of topologically protected wave modes in systems with properties that are modulated in space, and can be applied to the analysis of time-modulated media. The modulations are the results of projections from a higher dimensional space defining the strength of interactions. These projections may result in periodic, as shown herein, or quasi-periodic arrangements, as shown in related prior work. The analysis illustrates a mechanism for robust transfer of energy between two boundaries of a system employing elastic waves, which extends the abundant recent studies focusing on guiding of waves along interfaces separating two material phases of distinct topologies. The results provide guidelines for future designs of structural components or acoustic waveguides whose functionalities include the ability to selectively guide waves along desired paths, and to localize perturbation in predefined regions of the domains. 

\section*{Acknowledgments}
This work is supported by the National Science Foundation through the EFRI 1741685 grant.

\bibliographystyle{unsrt}
\bibliography{paper}

\end{document}